\documentclass[final,authoryear,5p,times,twocolumn]{elsarticle}

\usepackage{epsfig}

\usepackage{amssymb}
\usepackage{lineno}
\usepackage{rotating}
\usepackage{color,soul}
\usepackage{longtable}

\usepackage[bookmarks = true, bookmarksnumbered = true, pdfpagemode =None, pdfstartview = FitH, pdfpagelayout = SinglePage, colorlinks = true, urlcolor = blue, citecolor = monbleu]{hyperref}

\newcommand{\beq}{\bigskip\begin{equation}}
\newcommand{\eeq}{\bigskip\end{equation}}

\journal{Icarus}

\begin{document}

\begin{frontmatter}



\title{Jupiter's Para-H$_2$ Distribution from SOFIA/FORCAST and Voyager/IRIS 17-37 $\mu$m Spectroscopy}


\author[le]{Leigh N. Fletcher}
\ead{leigh.fletcher@leicester.ac.uk}
\author[idp]{I. de Pater}
\author[wtr]{W.T. Reach}
\author[idp]{M. Wong}
\author[jpl]{G.S. Orton}
\author[ox]{P.G.J. Irwin}
\author[rdg]{R.D. Gehrz}

%\author[om]{O. Mousis}
%\author[ox]{J.A. Sinclair}
%\author[umd]{R.K. Achterberg}
%\author[gsfc]{F.M. Flasar}

%\author[ox]{R.S. Giles}
%\author[ral]{J. Hurley}
%\author[cu]{N. Gorius}
%
%\author[umd]{B.E. Hesman}
%\author[gsfc]{G.L. Bjoraker}

\address[le]{Department of Physics \& Astronomy, University of Leicester, University Road, Leicester, LE1 7RH, UK}
\address[idp]{University of California, Berkeley, Astronomy Dept., 601 Campbell Hall, Berkeley, CA 94720-3411, USA}
\address[wtr]{Stratospheric Observatory for Infrared Astronomy, Universities Space Research Association, Mail Stop 232-11, NASA Ames Research Center, Moffett Field, CA 94035, USA.}
\address[jpl]{Jet Propulsion Laboratory, California Institute of Technology, 4800 Oak Grove Drive, Pasadena, CA, 91109, USA}
\address[ox]{Atmospheric, Oceanic \& Planetary Physics, Department of Physics, University of Oxford, Clarendon Laboratory, Parks Road, Oxford, OX1 3PU, UK}
\address[rdg]{Minnesota Institute for Astrophysics, School of Physics and Astronomy, 116 Church Street, S.E., University of Minnesota, Minneapolis, MN 55455, USA.}

%\linenumbers

\begin{abstract}

Spatially resolved maps of Jupiter's far-infrared 17-37 $\mu$m hydrogen-helium collision-induced spectrum were acquired by the FORCAST instrument on the Stratospheric Observatory for Infrared Astronomy (SOFIA) in May 2014. Spectral scans in two grisms covered the broad S(0) and S(1) absorption lines, in addition to contextual imaging in eight broad-band filters (5-37 $\mu$m) with spatial resolutions of 2-4".  The spectra were inverted to map the zonal-mean temperature and para-H$_2$ distribution ($f_p$, the fraction of the para spin isomer with respect to the ortho spin isomer) in Jupiter's upper troposphere (the 100-700 mbar range).  We compared these to a reanalysis of Voyager-1 and -2 IRIS spectra covering the same spectral range.  Tropospheric temperature contrasts match those identified by Voyager in 1979, within the limits of temporal variability consistent with previous investigations.  Para-H$_2$ increases from equator to pole, with low-$f_p$ air at the equator representing sub-equilibrium conditions (i.e., less para-H$_2$ than expected from thermal equilibration), and high-$f_p$ air and possible super-equilibrium at higher latitudes. In particular, we confirm the continued presence of a region of high-$f_p$ air at high northern latitudes discovered by Voyager/IRIS, and an asymmetry with generally higher $f_p$ in the north than in the south.  Far-IR aerosol opacity is not required to fit the data, but cannot be completely ruled out. We note that existing collision-induced absorption databases lack opacity from (H$_2$)$_2$ dimers, leading to under-prediction of the absorption near the S(0) and S(1) peaks.  There appears to be no spatial correlation between para-H$_2$ and tropospheric ammonia, phosphine and cloud opacity derived from Voyager/IRIS at mid-infrared wavelengths (7-15 $\mu$m).  We note, however, that para-H$_2$ tracks the similar latitudinal distribution of aerosols within Jupiter's upper tropospheric and stratospheric hazes observed in reflected sunlight, suggesting that catalysis of hydrogen equilibration within the hazes (and not the main clouds) may govern the equator-to-pole gradient, with conditions closer to equilibrium at higher latitudes. This gradient is superimposed onto smaller-scale variations associated with regional advection of para-H$_2$ at the equator and poles.

\end{abstract}

\begin{keyword}
Jupiter \sep Atmospheres, composition \sep Atmospheres, dynamics

\end{keyword}

\end{frontmatter}

%\linenumbers

%%%%%%%%%%%%%%%%%%%%%%%%%%%%%%%%%%%%%%%%%%%%%%
%%%%%%%%%%%%%%%%%%%%%%%%%%%%%%%%%%%%%%%%%%%%%%
%%%%%%%%%%%%%%%%%%%%%%%%%%%%%%%%%%%%%%%%%%%%%%
\section{Introduction}
\label{intro}

Far-infrared (IR) spectra of the giant planets are shaped by the collision-induced absorption of hydrogen and helium, providing a sensitive measure of the atmospheric temperature structure and the abundances of the most common gases in gas giant atmospheres \citep[e.g.,][]{79hanel, 80conrath, 83conrath, 98conrath, 00conrath}.  However, atmospheric studies in this spectral range are hampered by several factors.  Water vapour in Earth's atmosphere restricts our ability to measure the spectrum from the ground to narrow windows (known as the Q-band) between 17 and 24 $\mu$m.  Furthermore, diffraction-limited spatial resolutions worsen with increasing wavelength, meaning that observatories with large-diameter primary mirrors are required to resolve spatial contrasts.  Finally, it is difficult to disentangle the competing effects of a planet's temperature, aerosol distribution, helium abundance and the specific composition of hydrogen (para- or ortho-hydrogen) on the shape of far-IR spectrum.  To date, our only knowledge of the spatial variability of Jupiter's 17-55 $\mu$m spectrum comes from the Infrared Interferometer, Spectrometer and Radiometer (IRIS) experiments on the twin Voyager spacecraft \citep{74burke, 77hanel}.  Cassini did not have the spatial resolution to provide maps of Jupiter in this spectral range \citep{04flasar_jup}, and ISO could only provide disc-integrated views of the planet \citep{96kessler}.  In this study, we use the Faint Object infraRed CAmera for the SOFIA Telescope \citep[FORCAST,][]{10adams} on the Stratospheric Observatory for Infrared Astronomy \citep[SOFIA,][]{12young} to provide the first Earth-based spatially-resolved maps of Jupiter's full 17-37 $\mu$m spectrum for comparison with Voyager (previous ground-based efforts have been restricted to wavelengths shortward of $\sim24 \mu$m).  The maps allow us to assess Jupiter's tropospheric temperatures and the distribution of para-hydrogen as a tracer for atmospheric circulation.  Crucially, SOFIA flies at altitudes between 11-14 km, above 99\% of the Earth's water vapour, opening up the 17-37 $\mu$m region for infrared astronomy.

Voyager 1 and 2 flew by Jupiter in March and July 1979, respectively, providing spatially-resolved infrared spectra in the 180-2500 cm$^{-1}$ region (4-55 $\mu$m).  In the decades that followed, these data have been analysed by many authors to understand Jupiter's latitudinal temperature structure \citep{79hanel, 80conrath, 81conrath_stab, 83conrath, 84conrath, 86gierasch, 92carlson, 98conrath} and the thermal contrasts associated with discrete features like the Great Red Spot \citep{81conrath_grs, 81flasar, 92griffith, 96sada, 02simon, 06read_grs}.  Compositional studies focused on Jupiter's bulk helium abundance \citep{81gautier}, ammonia distribution \citep{86gierasch}, water ice signatures \citep{00simon} and hydrocarbon distributions \citep{10nixon}, highlighting the richness of the Voyager dataset.  

This work focusses on the subrange of Jupiter's far-infrared spectrum (270-600 cm$^{-1}$, 17-37 $\mu$m) that is accessible from SOFIA.  This range is dominated by the broad collision-induced H$_2$ S(0) and S(1) features near 354 cm$^{-1}$ (28.2 $\mu$m) and 587 cm$^{-1}$ (17.0 $\mu$m), respectively.  By measuring both features simultaneously, we can estimate the relative abundances of the two spin isomers of hydrogen - S(1) is formed from transitions within ortho-H$_2$ (the odd spin state of H$_2$ with parallel spins), whereas S(0) is formed from transitions within para-H$_2$ (the even spin state of H$_2$ with anti-parallel spins).  Populating or depopulating the S(0) states therefore affects the shape and gradient of the far-IR continuum in a way that is distinguishable from temperature changes, which tend to scale the spectral continuum brightness more uniformly over the entire wave band.   In `normal' hydrogen, at the high temperatures of the deep atmosphere, the two isomers should be in a 3:1 ratio \citep[a para-H$_2$ fraction $f_p$ of 0.25,][]{82massie}.  At the colder temperatures near the tropopause, the equilibrium $f_p$ increases to near 0.35, with lower temperatures favouring the increased population of the $J=0$ rotational state, the lower energy state of the S(0) transition.  However, vertical mixing of air parcels can cause significant disequilibrium, with regions of powerful uplift having lower para-H$_2$ fractions than the equilibrium (`sub-equilibrium' conditions) and regions of sinking beneath the tropopause that move high-$f_p$ air downwards (`super-equilibrium' conditions).  The rate of equilibration between the two gases is very slow \citep[multiple decades, although the influence of catalysis on aerosol surfaces remains unclear,][]{82massie}, such that fast mixing can significantly affect the distribution of para-H$_2$.   Changes in the relative populations of $J=0$ and $J=1$ rotational states therefore cause variations in the shape of the 270-600 cm$^{-1}$ spectrum, which can be used to trace vertical motions with Jupiter's troposphere. 

The distribution of Jupiter's para-H$_2$ measured by Voyager has been presented by multiple studies \citep{83conrath, 84conrath, 86gierasch, 92carlson, 98conrath}.  \citet{92carlson} used the similarity between the $f_p$ distribution and Jupiter's aerosol distribution to conclude that the spatial variations in $f_p$ were related to catalytic equilibration on aerosol particles.  Most recently, \citet{98conrath} used Voyager 1 spectra to map zonal-mean $f_p$ between 200-500 mbar, and showed that although the fine belt/zone structure was not well resolved, large-scale contrasts were evident.  Specifically, $f_p$ had a minimum at the equator and increased towards each pole, with the suggestion of an $f_p$ maximum over the north pole creating super-equilibrium conditions poleward of $45^\circ$N at the tropopause.  It is the existence of this hemispheric asymmetry in Jupiter's para-H$_2$ distribution, and the potential correlation with tropospheric aerosols, that we sought to test with the new SOFIA/FORCAST dataset.

Section \ref{obs} presents the FORCAST observations (both images and spectra) obtained in 2014.  The SOFIA spectra are modelled in Section \ref{models}, along with a re-analysis of Voyager/IRIS spectra to permit quantitative comparisons.  The resulting spatial distribution of $f_p$ is discussed in Section \ref{discuss} and compared to the latitudinal variability of tropospheric aerosols, ammonia and phosphine.  All latitudes in this paper are planetographic, all longitudes use System III west.

%%%%%%%%%%%%%%%%%%%%%%%%%%%%%%%%%%%%%%%%%%%%%%
%%%%%%%%%%%%%%%%%%%%%%%%%%%%%%%%%%%%%%%%%%%%%%
%%%%%%%%%%%%%%%%%%%%%%%%%%%%%%%%%%%%%%%%%%%%%%
\section{Observations}
\label{obs}

\subsection{SOFIA/FORCAST Observations}

The Faint Object infraRed CAmera for the SOFIA Telescope \citep[FORCAST,][]{10adams,12herter} is a thermal-infrared camera designed for the SOFIA airborne observatory \citep{12young, 09gehrz}.  It is comprised of two cameras (a short-wave camera covering 5-25 $\mu$m and a long-wave camera covering 25-40 $\mu$m) and grism spectroscopy.  The two arrays of FORCAST (a Si:As blocked impurity band (BIB) array for $\lambda<25$ $\mu$m and a Si:Sb BIB detector for $\lambda>25$ $\mu$m) operate at cryogenic temperatures (4K) and have a plate scale of 0.768 arcseconds per pixel.  The $256\times256$ array translates to a wide 191 arcsecond field of view that is more than sufficient to capture Jupiter's $\sim40$" disc.  When coupled with SOFIA's 2.7-m primary mirror (2.5-m effective aperture), this permits diffraction-limited observing for $\lambda>15$ $\mu$m, with an additional $1.3$" blurring due to the jitter of the telescope during flight.  The resulting angular resolution ranges from 2-4", depending on wavelength. The grism spectroscopy uses blazed diffraction gratings, and we specifically used the G227 (17.5-27.3 $\mu$m) and G329 (28.7-36.7 $\mu$m) grisms. 

\subsubsection{FORCAST Imaging}
\begin{figure*}
\begin{centering}
\centerline{\includegraphics[angle=0,scale=1.2]{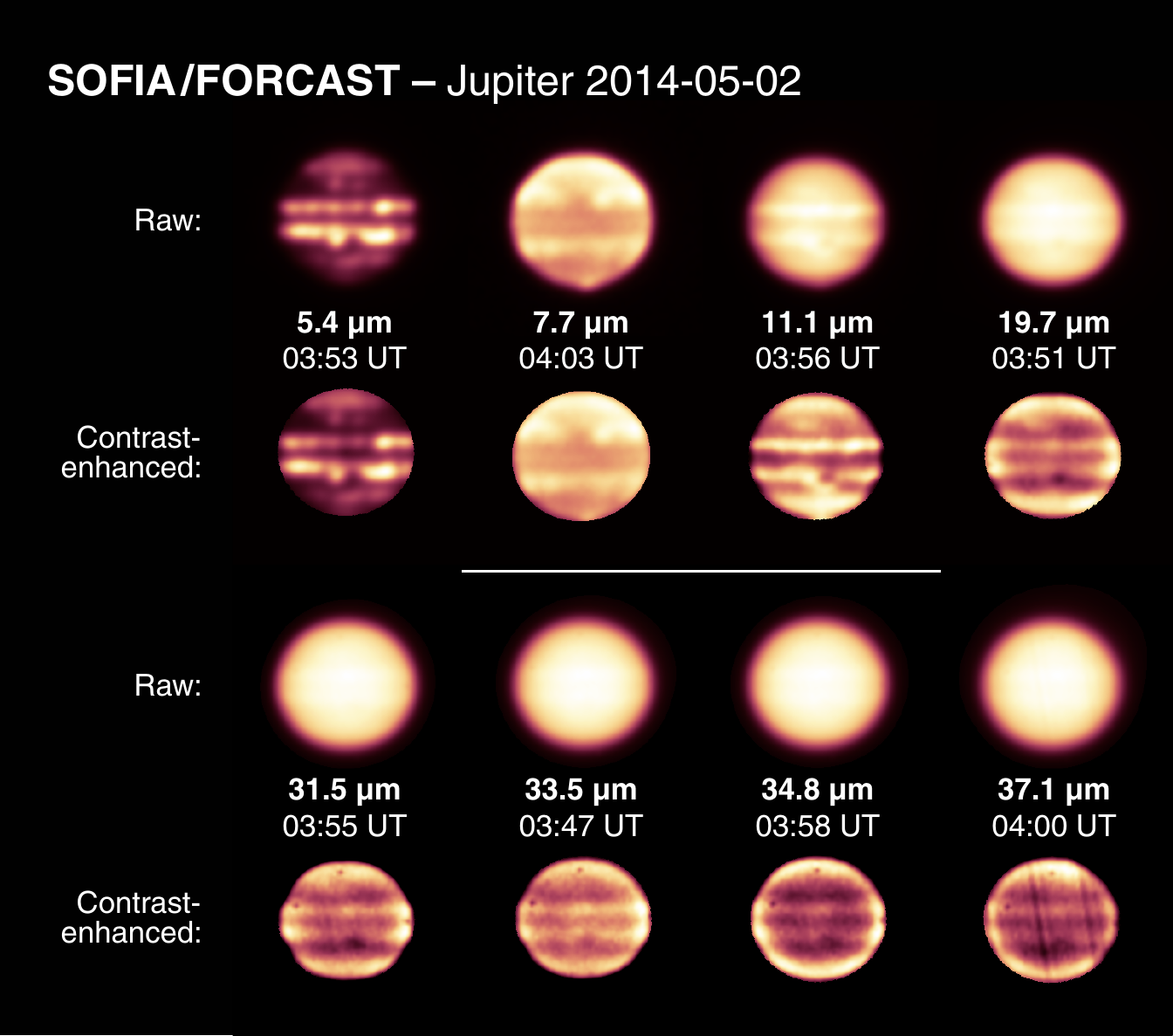}}
\caption{Contextual imaging obtained in FORCAST filters on May 2nd, 2014.  The Great Red Spot is in the centre of the disc.  Each filter is shown in the raw form and with an empirical limb-darkening correction applied to enhance the contrast, particularly at wavelengths beyond 30 $\mu$m.  This correction leads to some distortion in the oblate spheroidal shape of Jupiter at the edges. The central wavelengths and observation times are shown.  Artefacts due to pixel defects and readout striping affect the longest-wavelength images.    }
\label{images}
\end{centering}
\end{figure*}

Images were acquired in eight broad-band filters (Table \ref{tab:images}) in order to provide calibration and context for the spectroscopy observations in Section \ref{spec}.  Imaging in each filter was obtained in two epochs on May 2nd 2014: a first group between 03:42-04:03UT and a second group between 06:44-07:11UT (a planned third imaging set could not be acquired).  Examples of images acquired in the first group, with the Great Red Spot near the centre of the disc ($45^\circ$W longitude), are shown in Fig. \ref{images}.  For each filter, we compare the raw data (top row) with a contrast-enhanced version to show the spatial structure in the images (bottom row).  An empirical function was used to reproduce the emission angle dependence in each image ($\lambda>8$ $\mu$m) to correct for the limb darkening, without which very little structure can be seen in the raw images at $\lambda>30$ $\mu$m.  The true limb darkening of the planet is not resolved, so the apparent function is convolved with the point-spread function of the SOFIA primary mirror.  Furthermore, the long-wavelength images have contribution functions that sample the warm lower stratosphere at the highest emission angles, making simple Minneart-type corrections \citep{41minnaert} impossible.  

The absolute calibration of the images was determined by ensuring that photometric observations of standard stars, observed both on the flight containing the Jupiter observations and others in the flight series, matched the predicted flux within the FORCAST filters. The predicted fluxes were based on models for the spectral shape of the stellar emission, tied to absolute photometry from ground and space-based observations \citep{13herter}.  The absolute calibrators used for our images included $\alpha$ Boo, $\sigma$ Lib, and $\beta$ UMi.  For the 5.4--31.5 $\mu$m images, between 6 and 8 standard star observations were combined to determine the calibration factors. For the 37.1 $\mu$m image, 23 standard star observations were combined, to enhance the signal-to-noise at this wavelength.   The accuracy of the calibration for each filter is shown in Table \ref{tab:images}, with higher accuracy resulting from wider bandwidths (i.e., more photons in the wider filters), the usage of the filter (more usage means more calibrations are available), the type of calibrator (i.e., the brightness of the standard star) and the integration time.  For each filter, we summed the Jupiter flux to measure the spectral irradiance in Janskys, converting this to a disc-integrated brightness temperature using an effective Jupiter radius of 16.745" (i.e., accounting for the oblateness of the disc).  We repeated this for all longitudes observed (4-8 images per filter) to estimate the standard deviation, and show the median spectral irradiance and brightness temperatures in Table \ref{tab:images}.  These well-calibrated irradiance values will be used to cross-calibrate the FORCAST spectra in Section \ref{spec}.  

%The photometric precision of the imaging calibration (measured by the repeatability of the calibration star measurements corrected for observing conditions) is 1--3\%. 

\begin{figure*}
\begin{centering}
\centerline{\includegraphics[angle=0,scale=1.2]{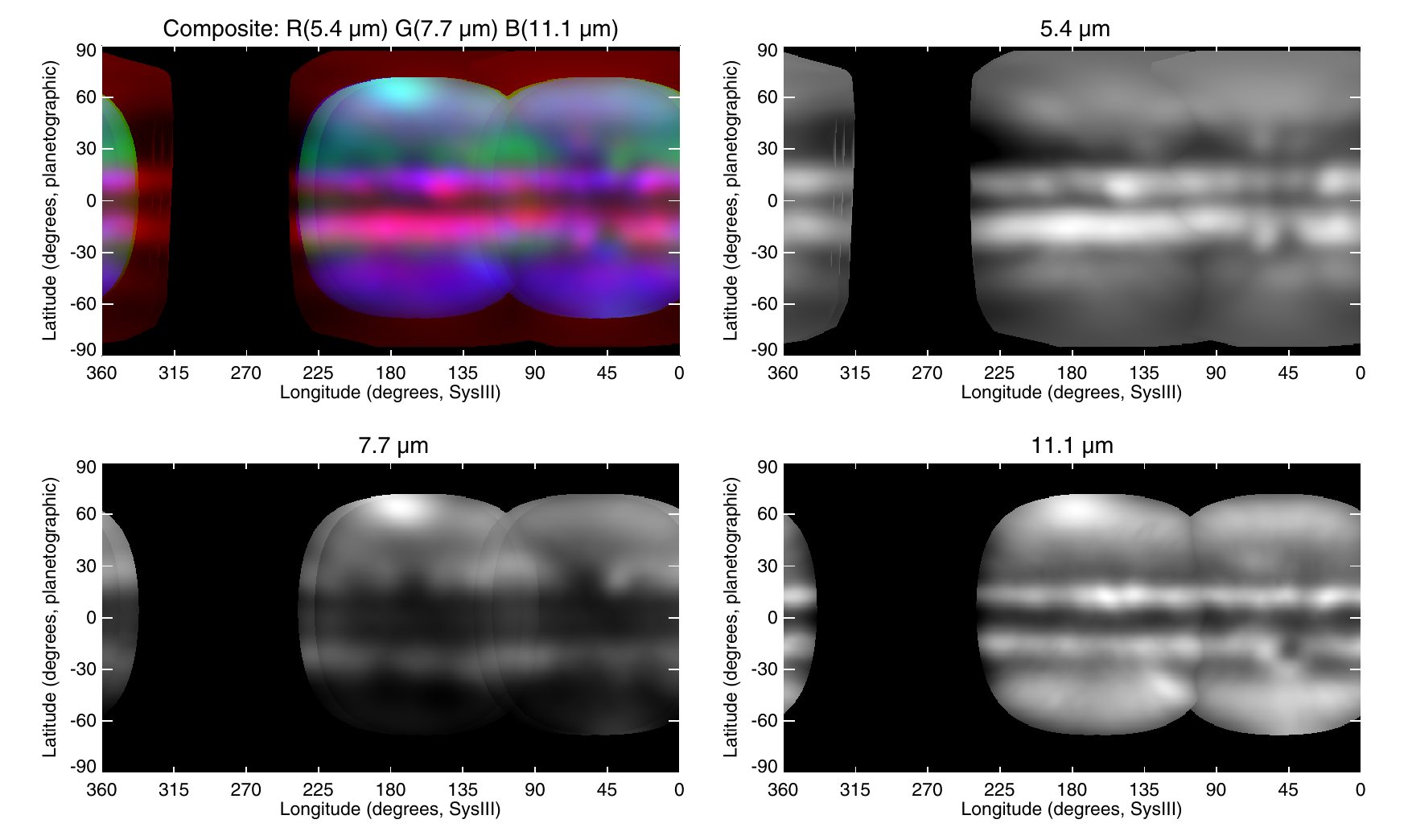}}
\caption{Cylindrical maps of multiple observations at 5.4, 7.7 and 11.1 $\mu$m.  5.4 $\mu$m samples radiance attenuated by Jupiter's 1-4 bar cloud opacity.  7.7 $\mu$m samples stratospheric methane emission near 10 mbar.  11.1 $\mu$m senses a combination of 500-mbar temperatures and NH$_3$ gas, although emission lines from stratospheric ethane dominate at higher emission angles.  The top left panel is an attempt to combine these three wavelengths into a crude 3-colour image to reveal the differences between the three filters. }
\label{cmaps}
\end{centering}
\end{figure*}

Images were obtained in two separate epochs to achieve near-global coverage, allowing us to create cylindrical maps in Fig. \ref{cmaps}.  These maps are limited to the M-band (5 $\mu$m) and N-band (7-13 $\mu$m) channels because longer wavelengths showed very little spatial contrast, with the exception of the cold equatorial zone (EZ) and warm neighbouring north and south equatorial belts (the NEB and SEB).  The cold Great Red Spot is visible in all images except 7.7 $\mu$m, where emission from CH$_4$ gas limits the sensitivity to higher, stratospheric altitudes near 10 mbar.  Longitudinal contrasts can be seen in Fig. \ref{cmaps}, primarily along the warm equatorial belts.  These cloud-free bands appear bright at 5.4 $\mu$m due to the negligible attenuation of radiance from the deeper troposphere, and also bright in the N-band near 11.1 $\mu$m due to warmer $500$-mbar temperatures in the belts compared to the zones.   There is a strong correlation between longitudinal wave patterns observed on the NEB and SEB in the M and N-band images, confirming that the dynamics modulating the tropospheric cloud opacity also modulate the temperature structure and ammonia humidity throughout the upper troposphere \citep[e.g.,][]{16fletcher_texes}.  At mid-latitudes we observe patches of brighter emission (particularly near $30^\circ$N, 45$^\circ$W) that are at the limit of SOFIA's spatial resolution.  These warm patches are potentially associated with small cyclonic ovals and brown barges that were visible in Jupiter's northern hemisphere in 2014.

Another notable aspect of the SOFIA maps in Fig. \ref{cmaps} is that we detect the warm polar hotspot near $180^\circ$W that is associated with upper stratospheric heating beneath the main auroral oval \citep[e.g.,][]{80caldwell, 93kostiuk, 93livengood}.  This bright emission shows up at 7.7 $\mu$m due to stratospheric methane, and near 11.1 $\mu$m due to stratospheric ethane. The 7.7-$\mu$m map shows a banded structure in the stratosphere consisting of a cool region equatorward of $\pm15^\circ$, as well as bands of bright emission between $\pm15-30^\circ$ in both hemispheres.  These stratospheric bands have been observed by Cassini \citep{04flasar_jup, 06simon} and ground-based observers \citep{94orton} and vary in their temperature over time.  Longitudinal contrasts in these warm stratospheric bands are associated with horizontal thermal wave activity at mid-latitudes, and such activity was prominent in 2014 \citep{16fletcher_texes}.  Finally, at the limit of the spatial resolution of SOFIA we observe a narrow, warm equatorial band at 7.7 $\mu$m, consistent with the local maximum in the thermal field associated with the current phase of Jupiter's Quasi-Quadrennial Oscillation \citep{91leovy,91orton}.  However, each of these M- and N-band spatial contrasts are better studied with larger telescopes on the ground.  Attempts to retrieve the thermal structure from these cylindrical maps \citep[following the imaging retrieval techniques of][]{09fletcher_imaging} met with limited success given that latitudinal contrasts in the uncorrected images in Fig. \ref{images} are so subtle.  Indeed, the main strength of the SOFIA/FORCAST dataset is not in the contextual images, but in the 17-37 $\mu$m spectroscopy offered by the grisms, as discussed in the following sections.

\begin{table*}[htdp]
\caption{Eight imaging filters used in this study, along with the disc-averaged brightness temperature and spectral irradiance measured in each filter. Uncertainties on the $T_B$ and total flux are the precision estimated from the standard deviation measured over multiple filters.  The large uncertainty at 5.4 $\mu$m is related to the large longitudinal variability observed at this wavelength.  The calibration uncertainty in the final column is the accuracy of determination of the fluxes of standard stars, based on repeatability of observations in a filter, after correction for atmospheric absorption, over multiple flights.}
%\makebox[\linewidth]{
\begin{center}
\begin{tabular}{|c|c|c|c|c|c|c|}
\hline
Filter & Wavenumber  &  Wavelength  & Bandwidth & Disc-Average $T_B$ & Total Flux & Calibration Unc.\\
name & (cm$^{-1}$) & ($\mu$m) & ($\mu$m) & (K) & (kJy) & (\%) \\
\hline
5.4 $\mu$m  &  1867.06  &   5.356  & 0.159  &  $198.0\pm29.1$  &    $6.9\pm1.0$ &  5.1 \\
7.7 $\mu$m  &  1297.86  &   7.705  & 0.465  &  $148.4\pm3.1$  &    $6.2\pm0.1$  &  4.2 \\
11.1 $\mu$m &   901.795 &  11.089  & 0.954  &  $129.1\pm4.9$  &   $26.2\pm1.0$  &  9.1 \\
19.7 $\mu$m &   507.305 &  19.712  & 5.506  &  $122.9\pm1.8$  &  $284.4\pm4.2$ &  2.3 \\
31.5 $\mu$m &   317.894 &  31.457  & 5.655  &  $122.8\pm3.7$ &  $654.0\pm20.0$ &  7.9 \\
33.5 $\mu$m &   298.507 &  33.500  & 5.658  &  $130.4\pm2.9$  &  $843.9\pm18.5$ & 12.7 \\
34.8 $\mu$m &   287.299 &  34.807  & 3.759  &  $130.5\pm4.9$  &  $857.8\pm32.2$ &  5.0 \\
37.1 $\mu$m &   269.222 &  37.144  & 3.284  &  $136.2\pm4.7$  &  $991.4\pm34.2$ &  9.9 \\
\hline
\end{tabular}
\end{center}
%}
\label{tab:images}
\noindent 
\end{table*}

\subsubsection{FORCAST Spectroscopy}
\label{spec}
Grism spectra were obtained with the $2.4\times191$ arcsecond slit (i.e., $\sim5$ times longer than Jupiter itself) and a spatial pixel scale of 0.768''.  Only the long-wave channel grisms were used in this study: G227 (17-27 $\mu$m) and G329 (28-37 $\mu$m) grisms to span 17-37 $\mu$m \citep{10keller}.  Data were obtained on May 3rd 2014, corresponding to a planetocentric solar longitude ($L_s$) of $158^\circ$, approaching the northern autumn equinox in February 2015.  At the time of observation, the slit (which has a fixed orientation in the instrument) was oriented approximately North-South on Jupiter.  For a single scan, twelve spectra were taken with each grism, starting at a position $-$20.5 pixels, i.e., $-$15.74'' from the centre of Jupiter, stepping across the planet from west to east in increments of 4 pixels, or 3.07''; the last scan was taken at a position $+$18.05''. The planet's diameter was 35.05'' at the time. Details describing the spectra are provided in Table \ref{tab:spectra}.  The spectra were taken using the standard IR observing two-position chop-with-nod, with a frequency of several Hz (Table \ref{tab:spectra}). The chop amplitude was $\pm30$'' and nod slews were 60''. 

The spectra, after pipeline processing through the step of spatial and spectral rectification (Rectified IMage, or RIM), were divided by the instrument response and atmospheric transmission curves; the latter was also used to determine the precise wavelength scale.  Individual spectra at each 3.07'' step along the scan were placed into a  3D cube, and then the cube was interpolated in the scan direction to generate square pixels, $0.768\times0.768$''. A polynomial fit to the background (on each row) was subtracted to remove residual flat-field variations and stray light.

\begin{figure}
\begin{centering}
\centerline{\includegraphics[angle=0,scale=0.6]{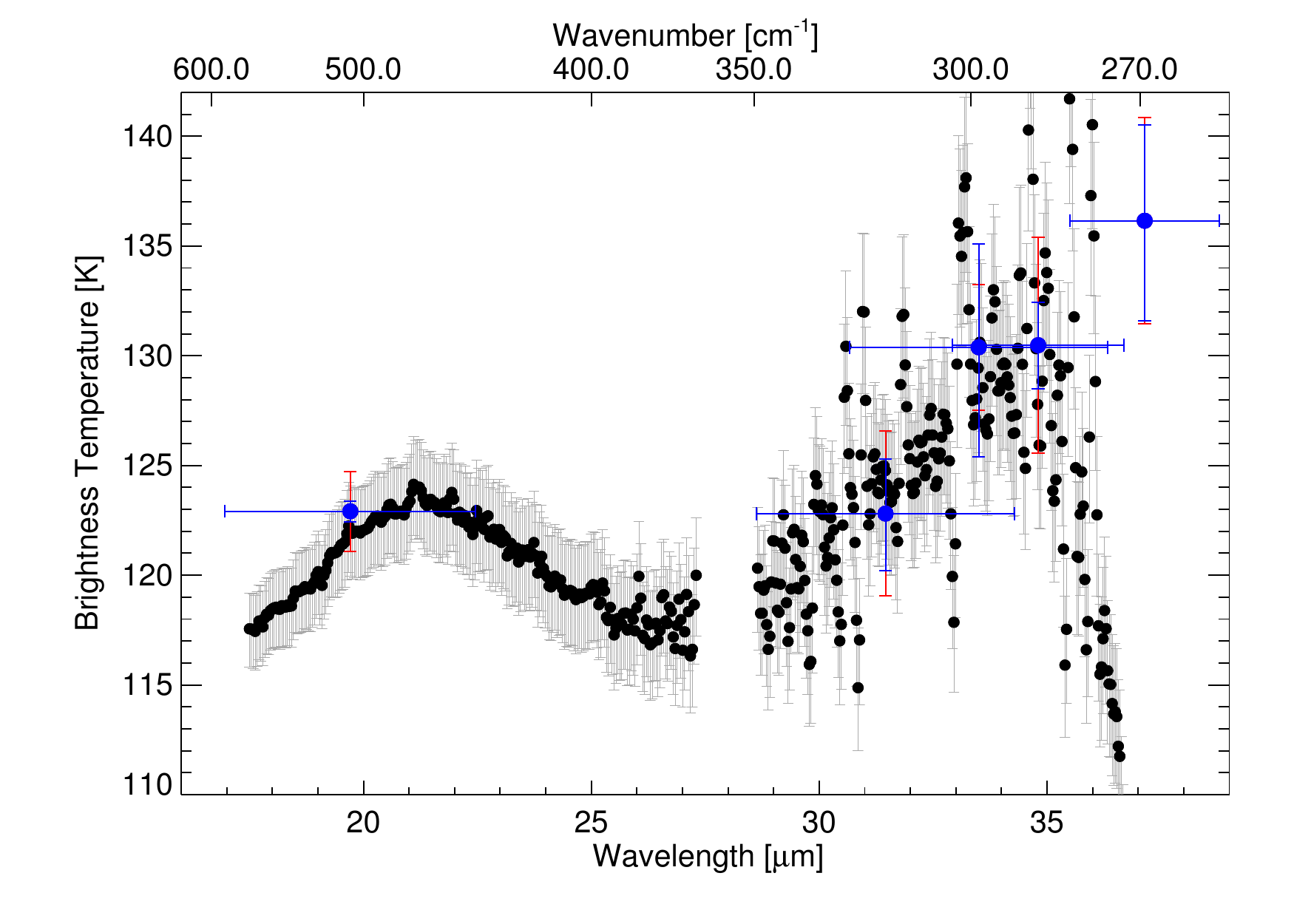}}
\caption{Comparison of the disc-integrated FORCAST spectra from the G227 (17-27 $\mu$m) and G329 (28-37 $\mu$m) grisms (black dots with grey error bars) with disc-integrated brightness temperatures derived from the images in Table \ref{tab:images}.  For each filter, we show the bandwidth (horizontal error bar), the standard deviation of the measurement from multiple images (red vertical error bar) and the absolute calibration uncertainty (blue vertical error bar).  The spectrum has been scaled to match the flux in the contextual images, and a 10\% radiometric uncertainty (converted to brightness temperature) is shown, representing the 2-12\% uncertainty range on the image calibration in Table \ref{tab:images}.  The downturn in the spectrum longward of 35 $\mu$m is a known issue with the G329 grism. }
\label{img-spx}
\end{centering}
\end{figure}

These final spectral data cubes were calibrated by summing the flux spatially over the disk of Jupiter and convolving spectrally over the bandpass of five of the broad-band filters (covering 17-37 $\mu$m), and scaling to match the flux of the broad-band images at the overlapping wavelengths.  Figure \ref{img-spx} compares the calibrated, disc-integrated spectrum to the fluxes in the five filters, indicating the increasing uncertainty with wavelength.  There was one spectral scan in the G227 grating, and three spectral scans in the G329 grating.  

\begin{figure*}
\begin{centering}
\centerline{\includegraphics[angle=0,scale=.80]{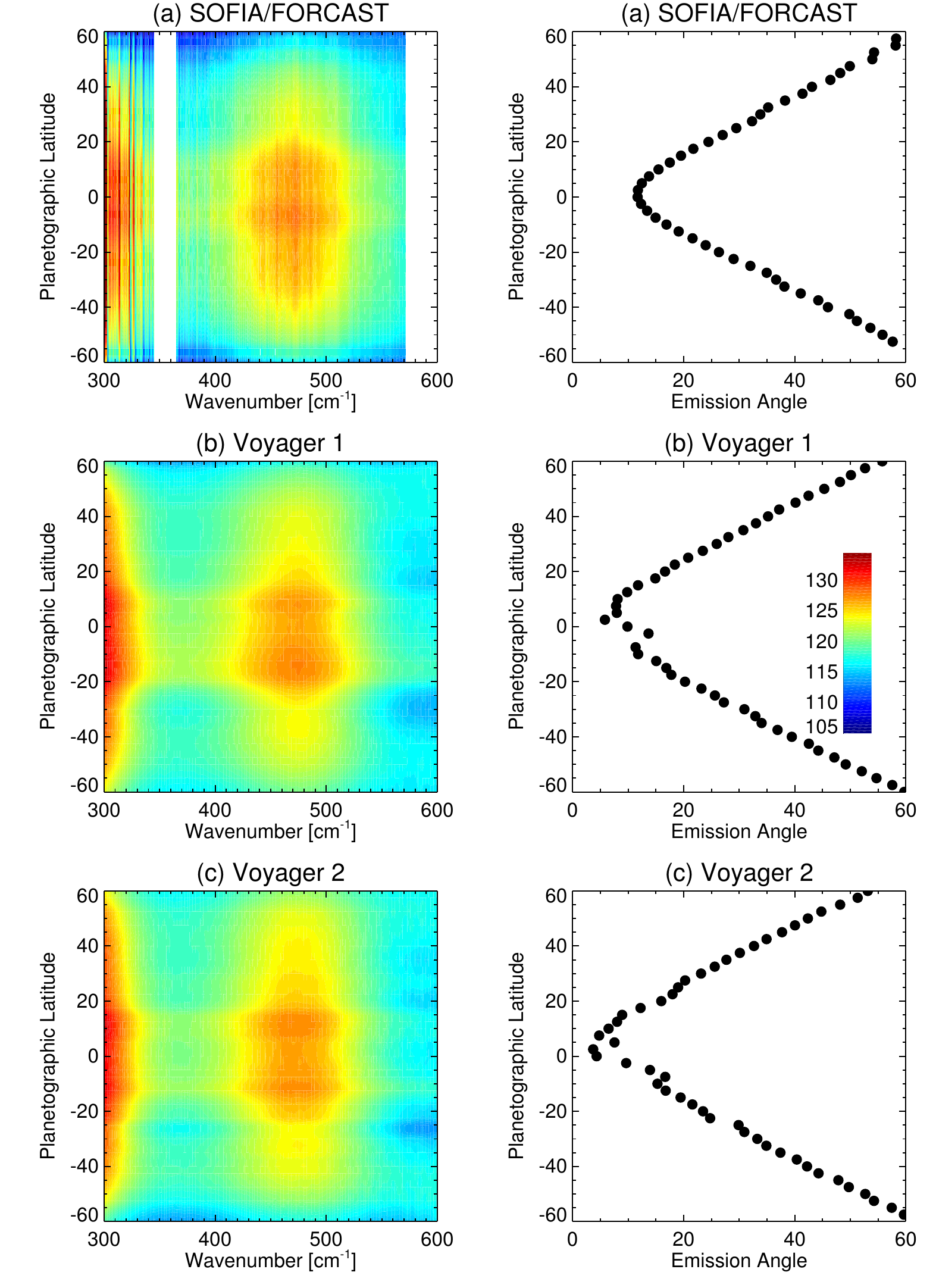}}
\caption{Comparison of the SOFIA/FORCAST and Voyager/IRIS brightness temperature spectra as a function of latitude and wavenumber.  The emission angles corresponding to each panel are shown in the right-hand column.  The colour-scale for the brightness temperatures is shown in the right-centre panel.  The white vertical band in the FORCAST spectrum in panel (a) is the gap between the two FORCAST grisms.}
\label{spxcontour}
\end{centering}
\end{figure*}

The measured flux in the spectra and images deviate substantially for $\lambda>33$ $\mu$m (Fig. \ref{img-spx}) due to known issues with the FORCAST G329 grism at the longest wavelengths.  The cause of this discrepancy remains unresolved, and we omit this range from our subsequent analysis.  Indeed, the fluxes measured in the context images are consistent with spectral models that suggest the radiance should increase for $\lambda>28$ $\mu$m towards the peak of Jupiter's black body emission beyond 50 $\mu$m.  The cubes were geometrically registered to determine the latitudes, longitudes and emissions angles corresponding to each pixel.  Zonal-mean spectra were generated by averaging pixels at each latitude within $20^\circ$ of the central meridian, using a $5^\circ$-wide latitudinal bin on a $2.5^\circ$ latitudinal grid between $60^\circ$N and $60^\circ$S.  These raw spectral data are shown in Fig. \ref{spxcontour}a, where we plot the radiance as a function of latitude and wavenumber and show how the emission angle for each spectrum varies from equator to pole.  These FORCAST spectra at specific latitudes are shown and modelled in Section \ref{models}.

\begin{table*}[htdp]
\caption{Details on the grism spectroscopy.}
%\makebox[\linewidth]{
\begin{center}
\begin{tabular}{|c|c|c|c|c|c|c|c|}
\hline
Grism & wavelength range & $R = \lambda/(\Delta \lambda)$ &UT time range & Longitude range & Latitude &Integration time & Chop freq. \\
\hline
G227 & 17.50-27.30 $\mu$m & 140 &05:36-05:45 & 265-270$^\circ$ &1.67$^\circ$&  5.66 s/scan & 3.97 Hz\\
G329 & 28.65-36.66 $\mu$m & 220 & 05:26-05:35 & 259-264$^\circ$ &1.67$^\circ$&  6.19 s/scan& 2.49 Hz\\
 \hline
\end{tabular}
\end{center}
%}
\label{tab:spectra}
\end{table*}%

 \subsection{Voyager/IRIS Observations}
 \label{voy_obs}
In order to perform a quantitative comparison of the temperatures and para-H$_2$ distributions from the FORCAST observations to previous data, we reanalysed spectra acquired by Voyager during the 1979 flybys.   The closest approaches (March 5, 1979 for Voyager 1 and July 9, 1979 for Voyager 2) correspond to planetocentric solar longitudes of $L_s=170.3^\circ$ and $L_s=180.4^\circ$, respectively, near the northern autumn equinox.  The Voyagers carried the Infrared Interferometer, Spectrometer and Radiometer (IRIS) instruments, with a Michelson interferometer providing 180-2500 cm$^{-1}$ spectra of the giant planets at a spectral resolution of 4.3 cm$^{-1}$ \citep{79hanel}.    IRIS spectra were extracted from the expanded volumes available on NASA's Planetary Data System\footnote{http://pds-rings.seti.org/voyager/iris/expanded\_volumes.html}.  Two temporal ranges were considered for each spacecraft, a dedicated north-south mapping sequence taken during the approach to Jupiter and a full average of all spectra acquired within 4 million km of closest approach (around $\pm72$ hours from closest approach).  The Voyager-1 inbound map was acquired from a distance of approximately 3 million km between 19:00 UT on March 2nd and 10:00UT on March 3rd 1979 \citep[see Fig. 1 of][]{86gierasch}.  This is the sequence that was used by \citet{83conrath, 98conrath} in their studies of Jupiter's para-H$_2$ and \citet{10nixon} in their study of Jupiter's hydrocarbons.  The Voyager-2 inbound map was acquired from a similar distance between 12:00UT on July 6th and 06:00UT on July 7th 1979.  At closest approach, the $0.25^\circ$ (4.4-mrad) diameter circular IRIS field of view provided spatial resolutions of 1230 km ($1.0^\circ$ latitude at the equator) and 2900 km ($2.3^\circ$ latitude at the equator) for Voyager 1 and Voyager 2, respectively.  In the larger average, these high-resolution observations are blended with spectra with resolutions as low as 17600 km ($14^\circ$ latitude at the equator) from 4 million km away.  

In all four cases we had to manually remove corrupted spectra that hadn't been caught by the `REJECT' flag in the database.  All datasets were binned on a $5^\circ$ latitude grid (with a $2.5^\circ$ step),  ensuring that the IRIS field of view was entirely on Jupiter's disc.  Only emission angles within $\pm10^\circ$ of the mean for each bin were retained.  Uncertainties were derived from the Voyager noise-equivalent spectral radiance (NESR), as described by \citet{14sinclair}.  We note that a misalignment of the Voyager-2 interferometer \citep{82hanel} led to slightly less sensitivity in the IRIS measurements in July 1979, and noisier spectra.  

The 17-33 $\mu$m (300-600 cm$^{-1}$) spectra from the two flybys are shown in Fig. \ref{spxcontour}b-c for comparison with the FORCAST data.  We are limited to this wavelength range by the poor quality and increasing noise in the long-wavelength FORCAST grism in the 270-300 cm$^{-1}$ region (not shown).   The broader wavenumber range for the IRIS data (300-1350 cm$^{-1}$) will be considered in Section \ref{discuss} to compare the retrieved para-H$_2$ distribution to zonal-mean distributions of ammonia, phosphine and clouds, following \citep{86gierasch}.  Fig. \ref{spxcontour} uses only the north-south mapping sequences, as these provided a smoother variation of emission angle with latitude.  The warmest brightness temperatures are seen at the lowest wavenumbers on the left in Fig. \ref{spxcontour}, but also near 470 cm$^{-1}$ in between the broad S(0) and S(1) features.  The three spectral panels show subtle asymmetries between the northern and southern hemispheres which appear to have changed with time - both FORCAST and Voyager-1 show the lowest brightness temperatures at high northern latitudes, whereas Voyager 2 shows the lowest brightness temperatures in the south, despite similar emission angles in the two Voyager sequences.  These asymmetries will be important when deriving the distribution of para-H$_2$ in Section \ref{models}.   

\begin{figure}
\begin{centering}
\centerline{\includegraphics[angle=0,scale=.90]{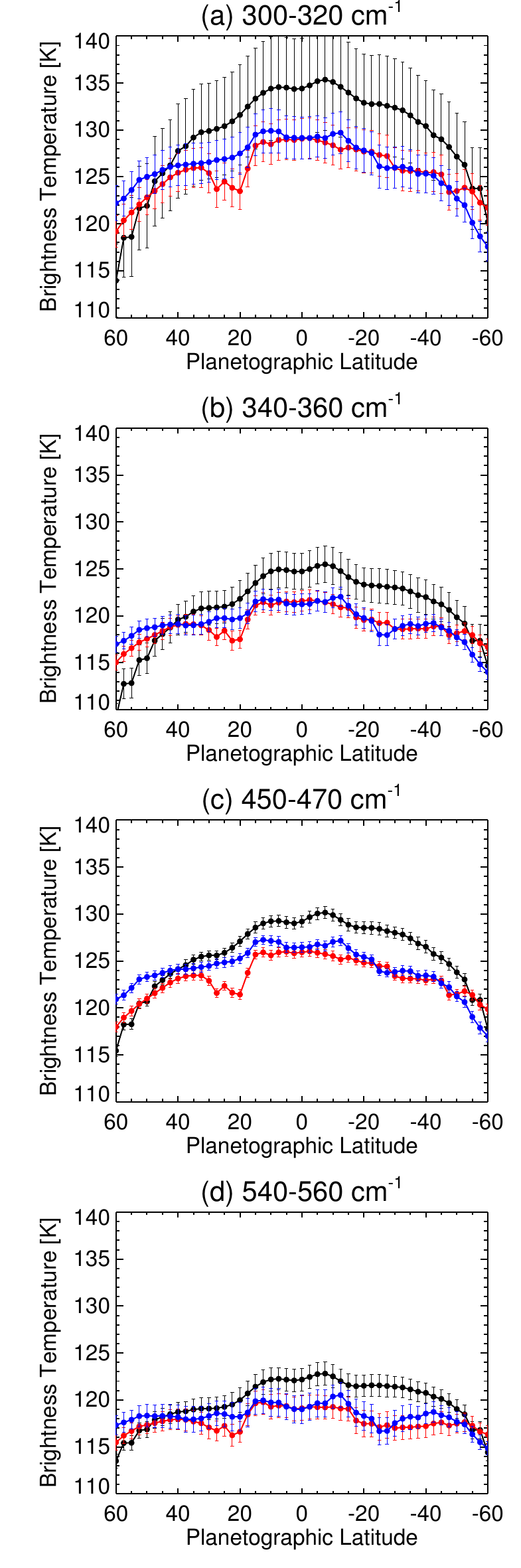}}
\caption{Comparison of zonal-mean brightness temperatures between SOFIA/FORCAST (black), Voyager 1 (red) and Voyager 2 (blue), averaged over the four spectral ranges shown.  The error bars show the standard deviation of the mean across the 20-cm$^{-1}$ wide bands, rather than the uncertainty of the coadded IRIS spectra, which is smaller and relatively uniform in this wavelength range.  A downward scaling of the SOFIA radiances by a factor of $\sim0.9$ makes them consistent with the Voyager radiances (see main text).}
\label{Tbcompare}
\end{centering}
\end{figure}

Finally, Fig. \ref{Tbcompare} shows a comparison between the FORCAST and IRIS brightness temperatures averaged in four spectral ranges coinciding with the lowest wavenumber portion of the FORCAST data, the S(0) and S(1) absorption centres and the peak emission in between.  The observations from the two Voyager encounters are consistent with one another, with the exception of cooler mid-northern latitudes during the Voyager 1 encounter.  The FORCAST data show similar spatial structure at low latitudes and an equator-to-pole drop due to limb darkening that will be accounted for in our spectral modelling.  The FORCAST data appear to be warmer than the IRIS data at all wavelengths, and we found that a $0.9\times$ scaling factor applied to the FORCAST spectra would bring the two into agreement.  This is within the 2-12\% calibration accuracy range in Table \ref{tab:images}.  We discuss the implications for this uncertainty in the radiometric calibration of FORCAST in Section \ref{models}.

%%%%%%%%%%%%%%%%%%%%%%%%%%%%%%%%%%%%%%%%%%%%%%
%%%%%%%%%%%%%%%%%%%%%%%%%%%%%%%%%%%%%%%%%%%%%%
%%%%%%%%%%%%%%%%%%%%%%%%%%%%%%%%%%%%%%%%%%%%%%
\section{Spectral Modelling}
\label{models}

\begin{figure}
\begin{centering}
\centerline{\includegraphics[angle=0,scale=0.7]{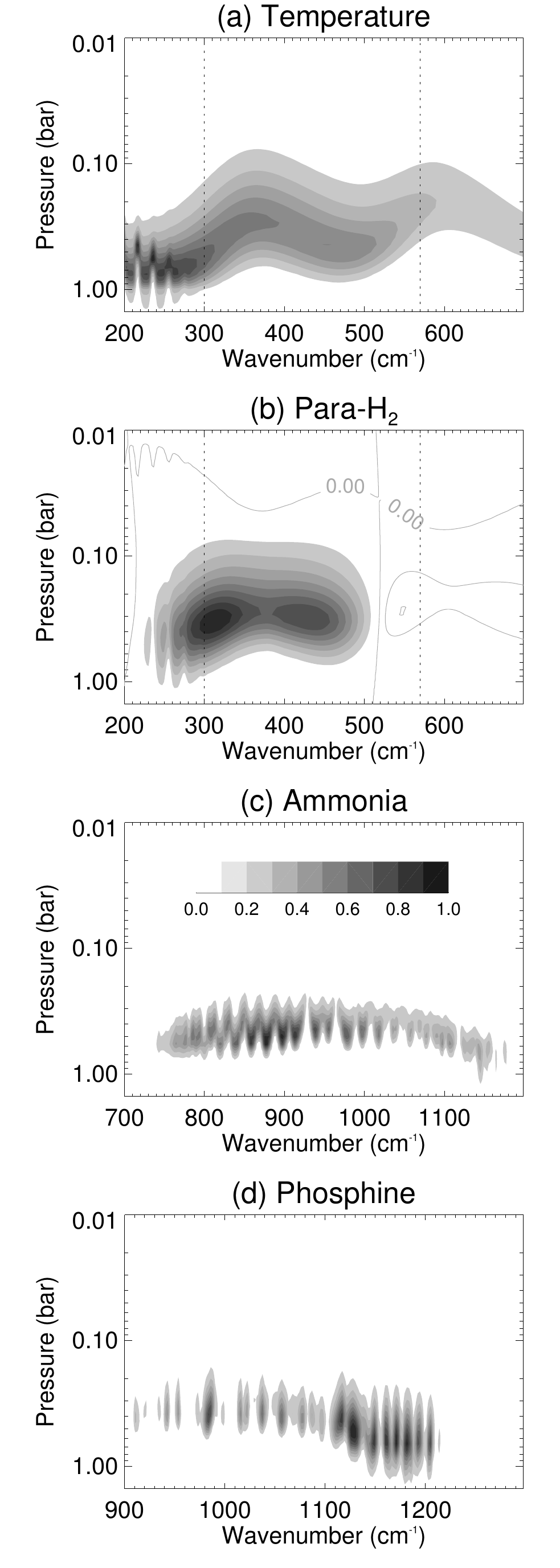}}
\caption{Jacobians (functional derivatives of radiance with respect to the parameter of interest) for tropospheric sensitivity in the far-IR (panels a,b) and mid-IR (panels c,d).  The Jacobians have been normalised to their maximum in this wavelength range, and contours are plotted in units of 0.1, as shown by the key in panel c.  In panel a, an \textit{increase} in temperature causes an \textit{increase} in the radiance (positive contributions). In panels b, c, and d, an \textit{increase} in the abundances of para-H$_2$, ammonia and phosphine cause a \textit{decrease} in radiance (negative contributions).   In panel b (para-H$_2$) we also show the zero line (grey line), as some weak positive contribution exists longward of 520 cm$^{-1}$.  Vertical dotted lines in panels a and b show the wavelength range of our SOFIA/FORCAST data.  The mid-IR influence of ammonia and phosphine will be discussed in Section \ref{discuss}.}
\label{jacobians}
\end{centering}
\end{figure}

SOFIA and Voyager spectra were analysed using the NEMESIS forward model and spectral inversion code \citep{08irwin}, which has been previously used to study the distribution of para-H$_2$ on Saturn \citep{16fletcher}, Uranus \citep{15orton} and Neptune \citep{14fletcher_nep}.  NEMESIS uses an optimal estimation retrieval architecture \citep{00rodgers} to maximise the quality of the fit to the data whilst remaining within the limits of physical plausibility using \textit{a priori} information.   The 17-33 $\mu$m spectrum is particularly simple, relying only on the collision-induced opacity of H$_2$-H$_2$ and H$_2$-He. Jacobians in Fig. \ref{jacobians}(functional derivatives of radiance with respect to the parameter of interest) demonstrate the vertical sensitivity of far-IR and mid-IR jovian spectra to tropospheric temperatures, para-H$_2$, ammonia and phosphine.  Spectra in this range are most sensitive to the temperature structure in the 100-700 mbar region, depending on the potential contribution of aerosol opacity \citep{98conrath}. There exists some additional sensitivity to the lower stratosphere (50-100 mbar) in the peaks of the S(0) and S(1) lines.  However, retrievals of temperature and para-H$_2$ become highly correlated at the low pressures near Jupiter's tropopause, and are likely to be unreliable in regions of aerosol opacity ($p>500$ mbar), so these variables can only be uniquely retrieved in the 200-500 mbar range, approximately \citep{98conrath}.

Our model atmosphere includes CH$_4$, NH$_3$, PH$_3$, C$_2$H$_2$ and C$_2$H$_6$, as described in \citet{09fletcher_ph3}.  NH$_3$ and PH$_3$ influence both mid-IR spectra (Fig. \ref{jacobians}c-d) and the far-IR spectra measured by Voyager (Fig. \ref{jacobians}a), so are varied in Section \ref{discuss} to compare their latitudinal distributions to that of para-H$_2$.  The prior $T(p)$ is based on a low-latitude average of Cassini infrared observations \citep{09fletcher_ph3}, which themselves used the $T(p)$ from the Galileo Atmospheric Structure Instrument \citep[ASI,][]{98seiff} as the prior.  The deep helium mole fraction was set to 0.136 based on \textit{in situ} Galileo probe measurements \citep{98niemann}, and we caution the reader that the absolute temperatures and para-H$_2$ fractions are sensitive to this assumption (although He is well-mixed, so this does not affect the measurement of relative spatial variability).  The prior for the para-H$_2$ fraction is based on equilibrium at the temperatures in our $T(p)$ profile, reaching a maximum of $f_p=0.34$ at the 113-K tropopause.  We retrieve continuous profiles of $T(p)$ and $f_p(p)$ simultaneously, whilst holding the abundances of all other gases constant at their priors.

\begin{figure*}
\begin{centering}
\centerline{\includegraphics[angle=0,scale=1.2]{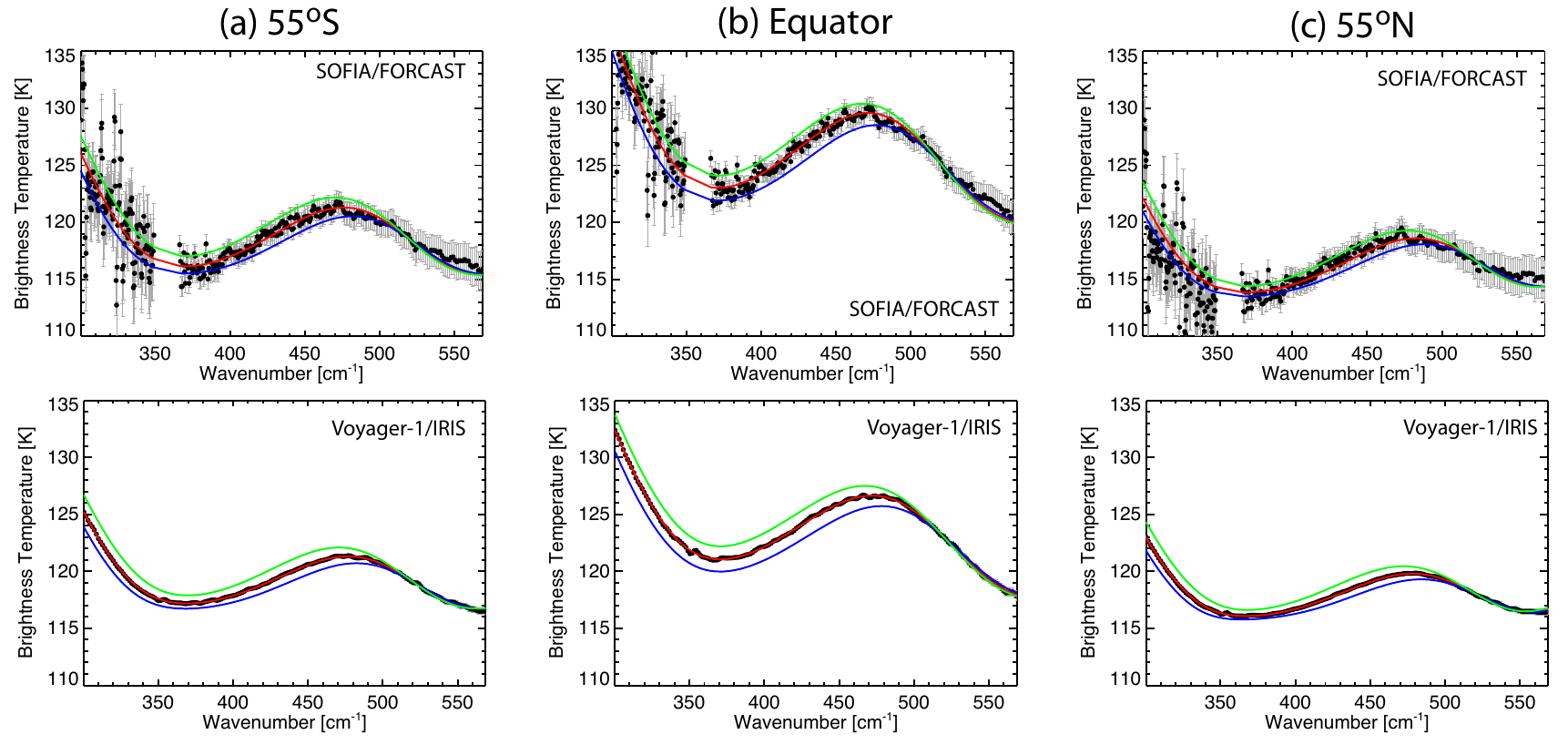}}
\caption{Best fits to the SOFIA/FORCAST (top row) and Voyager-1/IRIS (bottom row) 300-570 cm$^{-1}$ spectra at three locations: $55^\circ$S, the equator and $55^\circ$N.  Spectra were coadded within $\pm2.5^\circ$ of these central latitudes, and the small uncertainties associated with the IRIS spectra cannot be seen at this scale.  Brightness temperature scales are the same for all the figures to show differences, although the primary effect is limb darkening from the equator to the high latitudes. The red lines show the spectral model produced with our best-fitting $T(p)$ and $f_p(p)$.  The blue/green lines show the sensitivity of the spectra to uniformly increasing/decreasing the para-H$_2$ fraction by 0.05, the upper limit of the latitudinal variability of the para-H$_2$ fraction detected on Jupiter.}
\label{specfits}
\end{centering}
\end{figure*}

Fig. \ref{specfits} compares the FORCAST and IRIS data to our best-fitting spectral models at three latitudes.  The sensitivity of the spectra to changes in para-H$_2$ is shown by offsetting the $f_p$ distribution by $\pm0.05$ at all altitudes, demonstrating that the shape of the 300-500 cm$^{-1}$ (centred on the S(0) line) is sufficiently altered, even at the highest latitudes and emission angles, to allow us to derive an $f_p$ distribution from these data. Furthermore, as the para-H$_2$ change affects a broad swathe of the continuum, the high spectral sampling of multiple independent measurements allows us to derive a best-fit $f_p$ with uncertainties of $\approx±0.02$.  This figure also confirms the difference between the spectra at $55^\circ$N and $55^\circ$S, which has implications for the para-H$_2$ asymmetry discussed below.  The following subsections describe the challenges associated with analysing these data.

\subsection{Collision-induced absorption model}

\begin{figure}
\begin{centering}
\centerline{\includegraphics[angle=0,scale=.70]{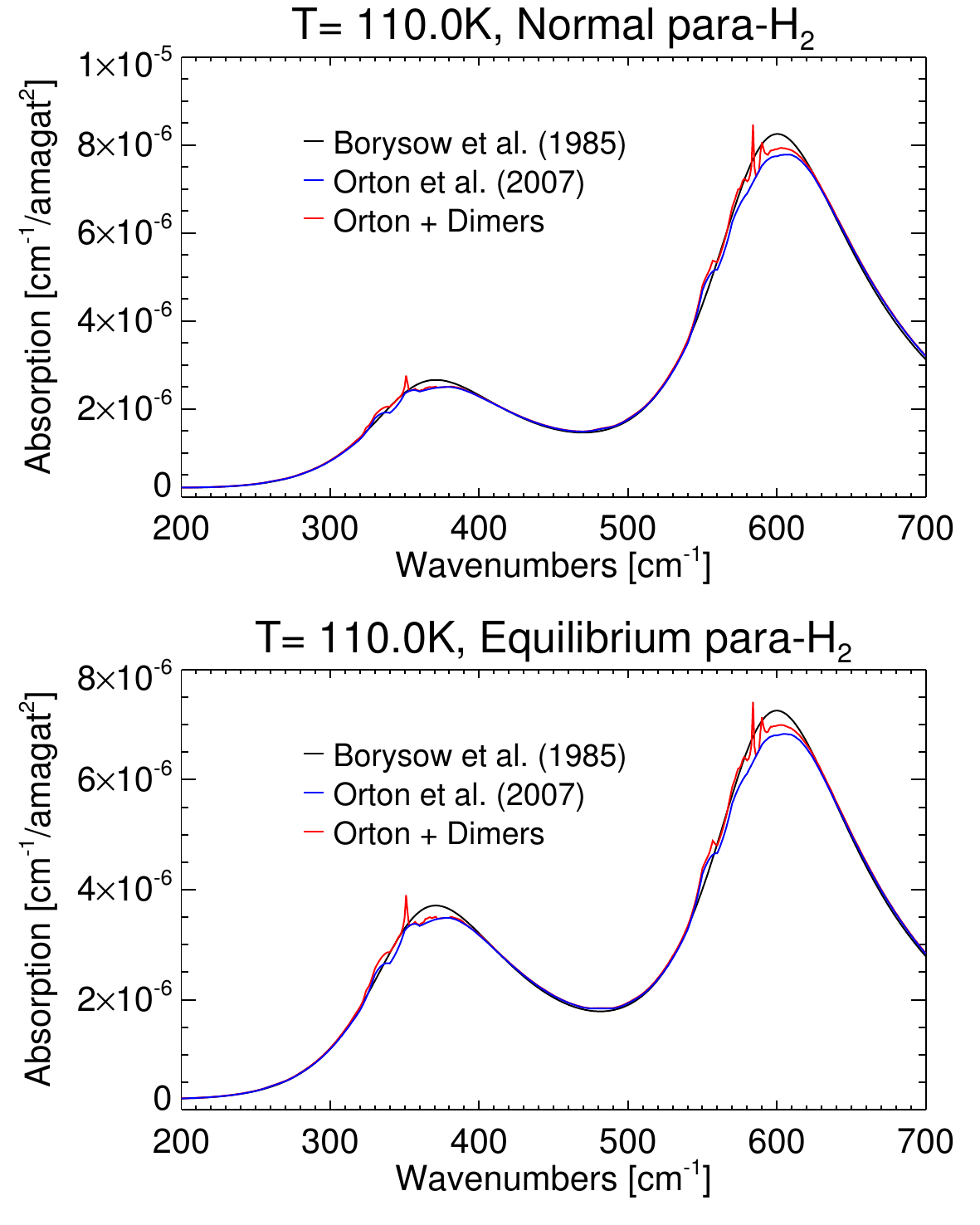}}
\caption{H$_2$-H$_2$ absorption coefficient, normalised by the square of the gas density, at 110 K for `normal' hydrogen (para-H$_2$ fractions of 0.25, the high-temperature asymptote) and equilibrium H$_2$ at 110 K, representative of the coldest temperatures near Jupiter's tropopause).  The black line is for the model of \citet{85borysow}, the blue line is for \citet{07orton} \citep[identical to that included in HITRAN 2012,][]{12richard}. The red line shows the increased opacity from adding (H$_2$)$_2$ dimer absorption to the CIA from \citet{07orton}.}  
\label{cia}
\end{centering}
\end{figure}

Two different calculations of the collision-induced H$_2$-H$_2$ opacity are available for our forward model - \citet{85borysow} provided semi-empirical functions based on \textit{ab initio} studies of H$_2$-H$_2$ collisions over a broad spectral range; \citet{07orton} then revised these calculations using updated quantum line shape calculations.  Both used \textit{ab initio} dipole data from \citet{89meyer,89meyer_fund} with a stated accuracy of 3\%, and neither incorporated the presence of (H$_2$)$_2$ dimers near to the S(0) and S(1) lines.  However, \citet{07orton} found that some of the dipole components of \citet{85borysow} had been overestimated due to an accounting error in the original work.  The compilation of \citet{07orton} now forms the `Alternative' CIA database maintained by HITRAN \citep{12richard}.  While the authors stated that differences were found for wavenumbers beyond 600 cm$^{-1}$ and temperatures below 120 K, we find that the two collision-induced-absorption (CIA) resources predict different opacities in the S(0) and S(1) lines at temperatures relevant to Jupiter's atmosphere.  Fig. \ref{cia} compares the calculations of \citet{85borysow} and \citet{07orton} at 110 K (jovian tropopause temperatures) and two different para-H$_2$ fractions.  At equilibrium, the \citet{85borysow} model predicts around 7\% higher absorption in the S(0) and S(1) peaks than the model presented by \citet{07orton}. This has implications for our fits to the 17-33 $\mu$m spectra.

\begin{figure*}
\begin{centering}
\centerline{\includegraphics[angle=0,scale=.80]{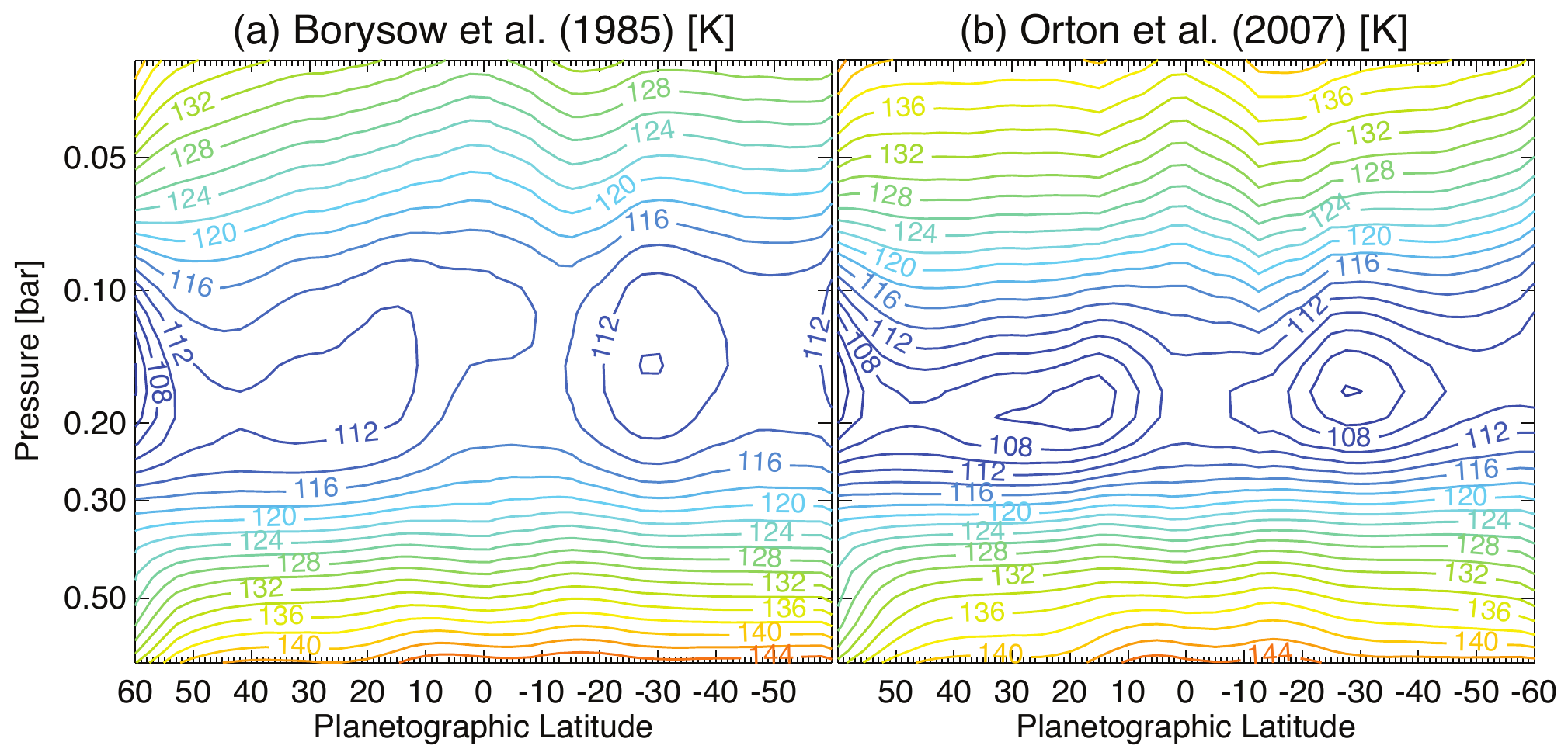}}
\caption{Comparison of zonal mean temperature cross-sections derived from the Voyager-1 north-south mapping sequence assuming CIA opacities from (a) \citet{85borysow} and (b) \citet{07orton}. Temperature precision ranges from 2 K at 700 mbar to 1.2 K at 400 mbar and 1.8 K at the tropopause, but this does not encompass the systematic differences between the two sources of CIA. }  
\label{cia_comp}
\end{centering}
\end{figure*}

Indeed, when we run our inversion of Voyager-1/IRIS spectra using the two different opacity databases we derive wholly different zonal mean temperature cross-sections (see Fig. \ref{cia_comp}).  Fits with the CIA opacity from \citet{85borysow} are qualitatively similar to those presented in previous studies \citep[e.g.,][]{98conrath}, with a broad tropopause region between 100-200 mbar.  Conversely, fits with the CIA opacity from \citet{07orton} require a much narrower tropopause region near 200 mbar and warmer temperatures in the lower stratosphere.  Given that the latter CIA database has a smaller contrast between the absorption maxima at 354 and 587 cm$^{-1}$ and the minimum near 460 cm$^{-1}$, our inversion requires warmer lower-stratospheric temperatures to reproduce the brightness temperatures near the peaks of the S(0) and S(1) features.  This is counter to the findings of all previous inversions of IRIS data, but cannot be discounted given that the newer calculations of \citet{07orton} are considered to be an improvement on those of \citet{85borysow}.  Furthermore, the newer calculations showed better agreement with the experimental results of \citet{96birnbaum}.  Nevertheless, the newer calculations generally produce poorer fits to the data (worsening the $\chi^2/N$ by a factor of two at all latitudes).  

We suggest that additional (H$_2$)$_2$ dimer absorption \citep[a molecule bound by van der Waals forces, ][]{84frommhold,90schaefer,96birnbaum} is required to resolve this discrepancy.  In their study of Uranus' H$_2$-He spectrum, \citet{14orton} incorporated new calculations of the dimer opacity (provided by J. Schaefer) to their existing CIA database \citep{07orton}.  A $\sim5$\% enhancement in the absorption near the S(1) peak is shown as the red curve in Fig. \ref{cia}, bringing the total absorption closer to the original erroneous values of \citet{85borysow}.  Furthermore, \citet{84frommhold} discussed dimer features near the S(0) line in Voyager/IRIS spectra of Jupiter, and found that bound-free transitions between dimers contributed approximately 5\% to the opacity at 120 K. However, the dependence of the dimer contribution on the para-H$_2$ fraction has not been modelled, and inclusion of these new calculations produced features in our CIA database that are not directly observed, implying that there is room for improvement of the theoretical simulations.   We therefore use the CIA opacity from \citet{85borysow} in this paper for simplicity and comparison to previous works, and caution the reader about the `missing absorption' in the HITRAN compilation \citep{07orton, 12richard} near the peaks of the S(0) and S(1) lines.

\subsection{Aerosol influence and spectral windowing}

The reliable portion of the SOFIA/FORCAST data covers the 300-570 cm$^{-1}$ (17.5-33.3 $\mu$m) spectral range.  In their study of Voyager/IRIS spectra, \citet{98conrath} noted that spectra below 320 cm$^{-1}$ and between 430-520 cm$^{-1}$ needed to be excluded due to the possible presence of aerosol opacity \citep{92carlson}.  \citet{92carlson} showed that the inclusion of opacity from large aerosols (with radii comparable to the wavelength) would shift the Jacobian in the absorption minimum (near 460 cm$^{-1}$ in Fig. \ref{jacobians}a, 21.7 $\mu$m) upwards to cooler altitudes, thereby reducing the overall contrast in the 300-570 cm$^{-1}$ spectrum.  Subsequent retrieval studies have utilised compact aerosol layers with a base near 800 mbar \citep[consistent with the expected condensation altitude of a $3\times$ solar enriched NH$_3$ gas,][]{99atreya} to provide thermal-infrared opacity \citep{04wong, 05matcheva, 06achterberg, 09fletcher_ph3}.  Using the Voyager-1 north-south mapping sequence, we re-ran the zonal mean temperature and para-H$_2$ inversion using three different assumptions: (1) zero aerosol opacity; (2) a globally-homogeneous cloud layer with fixed optical depth of unity at 1 bar; and (3) allowing the aerosol opacity to vary as a free parameter during fits.   The cloud was included as a compact cloud (aerosol-to-gas scale height ratio of 0.2) with a base at 800 mbar.  Each experiment was repeated twice, once with the full 300-570 cm$^{-1}$ range and once restricting to the 320-430 cm$^{-1}$ and 520-570 cm$^{-1}$ ranges following \citet{98conrath}.

\begin{figure*}
\begin{centering}
\centerline{\includegraphics[angle=0,scale=.80]{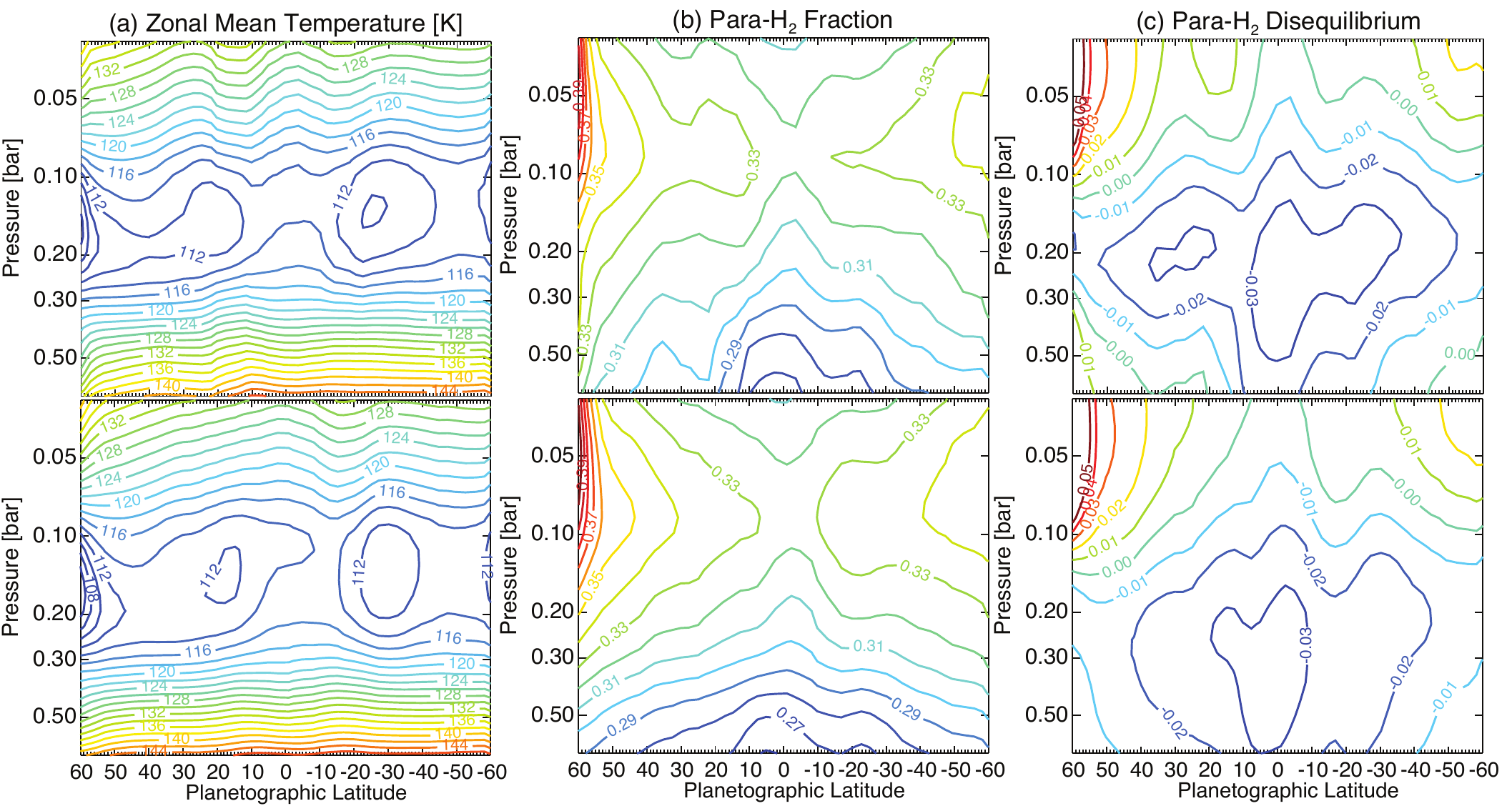}}
\caption{Comparison of the temperature, para-H$_2$ and para-H$_2$ disequilibrium retrievals without (top row) and with (bottom row) the inclusion of aerosol opacity.  Results are shown for fits of the full 300-570 cm$^{-1}$ range of the Voyager-1/IRIS north-south mapping sequence.  Temperature uncertainties range from 2 K at 700 mbar to 1.2 K at 400 mbar and 1.8 K at the tropopause; $f_p$ uncertainties range from 0.02 at 600 mbar and 100 mbar to 0.01 at 200 mbar.}
\label{aer}
\end{centering}
\end{figure*}

$T(p)$, $f_p$ and para-H$_2$ disequilibrium retrievals with and without aerosols are compared in Fig. \ref{aer}.  When fitting the full spectral range, we found closer fits to the data when no aerosol opacity was included (Fig. \ref{aer} top row), and negligible difference between the fits when either homogeneous or inhomogeneous cloud layers were used (Fig. \ref{aer} bottom row).  In this instance, this argues against the addition of aerosol opacity of any kind to the 300-570 cm$^{-1}$ fits.  Furthermore, when aerosols were allowed to vary we obtained only a geometrical effect (a correlation with the emission angle), with lower optical depths required at the equator ($\sim0.3$) and higher optical depths near the poles ($>2$).  This suggests that our assumed compact vertical distribution of aerosol opacity is incorrect.  Furthermore, the goodness-of-fit was substantially improved at the equator when no aerosols were included, precisely in the region that would be expected to show the highest opacity.  Unfortunately, neither the low-resolution FORCAST data nor the IRIS data provide the necessary centre-to-limb observations to further constrain this vertical distribution.  The presence or absence of aerosols therefore adds uncertainty to our inversions - the addition of aerosols permits a warmer atmosphere at $\sim700$ mbar by 1-2 K and a slightly cooler atmosphere near 300 mbar by $<1$ K, and negligible differences for $p<300$ mbar.  The addition of aerosols causes systematic changes to $f_p$ of around 0.01, and these are most apparent for $p>500$ mbar where $f_p$ can no longer be constrained.  

Finally, we find that omission of the 300-320 cm$^{-1}$ and 430-520 cm$^{-1}$ regions has negligible impact on our retrieved $T(p)$ and $f(p)$ distributions.  Changes to the $f_p$ values are smaller than 0.01 and changes to the tropospheric temperatures are smaller than 1 K, both within the formal uncertainties on the $T(p)$ and $f_p(p)$ structures shown in Fig. \ref{aer}.  Crucially, the latitudinal contrasts in temperature and $f_p$ are the same irrespective of how we include aerosols and spectral windowing in the inversion, and these contrasts match those of \citet{98conrath} extremely well.  We will return to the importance of aerosols in Section \ref{discuss}.

\subsection{Spectral selection from Voyager}

\begin{figure*}
\begin{centering}
\centerline{\includegraphics[angle=0,scale=.90]{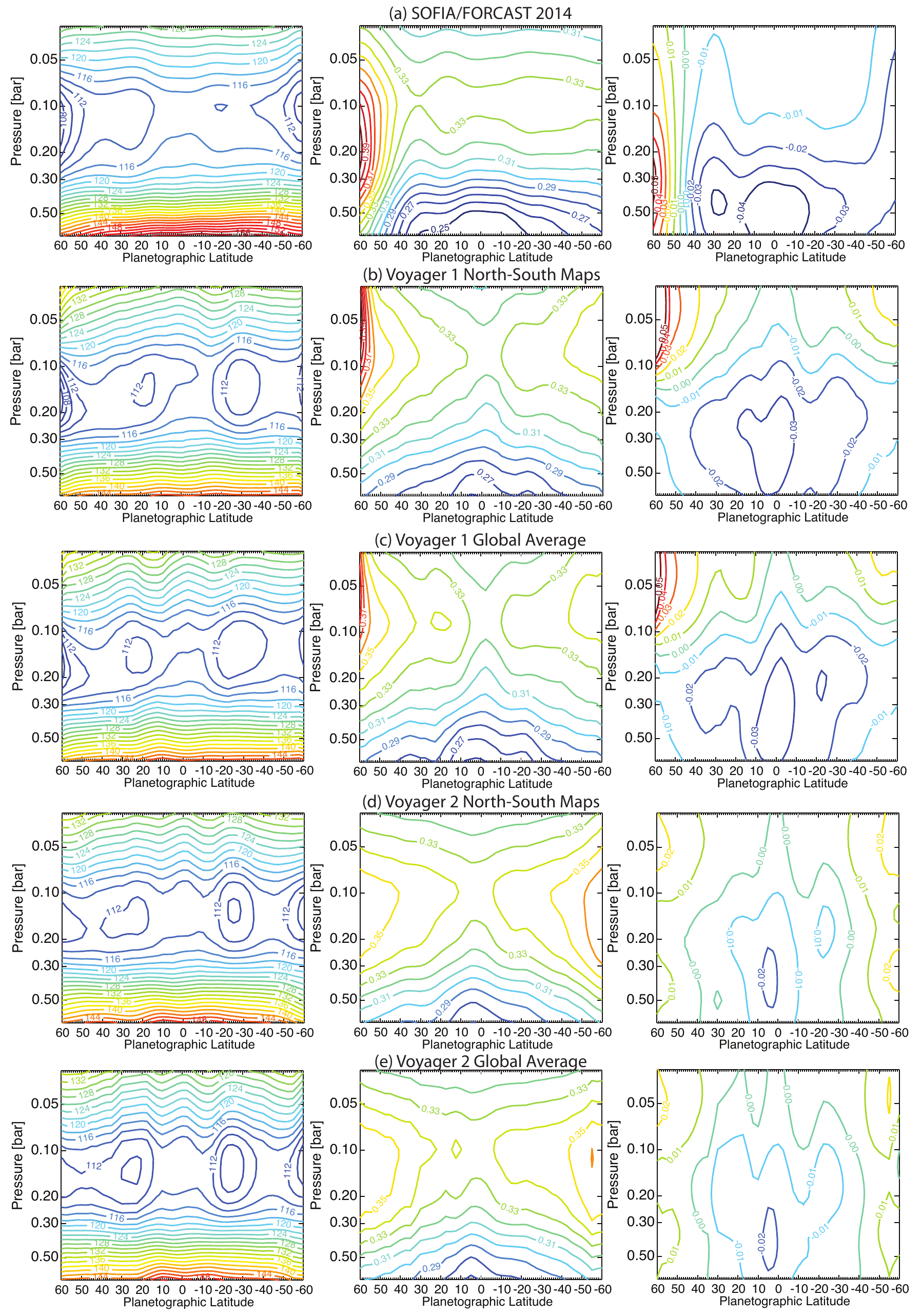}}
\caption{Comparison of the retrieved atmospheric structure from four different Voyager spectral averages and SOFIA/FORCAST.  The left-hand column shows the $T(p)$ retrieval with 2-K contours, with uncertainties ranging from 2 K at 700 mbar to 1.2 K at 400 mbar and 1.8 K at the tropopause.  The centre column shows the $f_p(p)$ retrieval with 0.01 spacing, with uncertainties ranging from 0.02 at 600 mbar and 100 mbar to 0.01 at 200 mbar.  The right-hand column shows the degree of disequilibrium with 0.01 contours, from sub-equilibrium at low-latitudes to super-equilibrium at high latitudes.}
\label{comp_all}
\end{centering}
\end{figure*}

The tests conducted so far used only the north-south mapping sequence from Voyager 1.  However, Section \ref{voy_obs} describes the coaddition of spectra to create `global averages' of each Voyager dataset within 4 million km of Jupiter and one north-south mapping sequence for each spacecraft.  These four different datasets are compared to the zonal mean $T(p)$, $f_p(p)$ and para-H$_2$ disequilibrium from SOFIA/FORCAST in Fig. \ref{comp_all}.  Similar latitudinal structures are evident and will be discussed in Section \ref{discuss}.  However, we note that there are differences depending on (i) whether we just take the north-south maps or include the full Voyager dataset; and (ii) whether we consider Voyager 1 or 2, only 4 months apart.  In the first case, differences are due to the blending of a range of spatial resolutions in the global averages and the higher signal-to-noise of the coadds (more spectra will reduce the noise and weight the inversion more strongly towards the data).  In the second case, the Voyager-2 spectra were generally noisier due to the misalignment of the Voyager-2 interferometer \citep{82hanel} discussed in Section \ref{voy_obs}, which results in the retrieval being more strongly weighted to the prior.  However, two features of the para-H$_2$ distribution are robust no matter how we process the dataset - there is a minimum in $f_p$ (and hence sub-equilibrium conditions) at the equator; and an asymmetry in $f_p$ between the northern and southern high latitudes which will be discussed in Section \ref{discuss}.

Close inspection of the fits to the Voyager 1 and Voyager 2 spectral averages revealed that the residual always took on a common undulating shape, indicating that we were underfitting one side of the $\sim470$-cm$^{-1}$ peak and overfitting the other.  This undulation in the residual was worse for Voyager 2 than for Voyager 1, potentially as a result of the interferometer mis-alignment.  To investigate the possibility of a wavelength offset in the IRIS spectra, we reran the zonal-mean retrievals shifting the spectrum in 0.25-cm$^{-1}$ steps between 0 and 4.0 cm$^{-1}$ (i.e., one FWHM for IRIS) using the CIA of \citet{85borysow}.  Better fits to the data could be obtained with shifts of 0.4-1.1 cm$^{-1}$ for Voyager 1 and 1.6-2.5 cm$^{-1}$ for Voyager 2.  However, such large shifts are not viable, given that NH$_3$ rotational features in the 200-250 cm$^{-1}$ of the IRIS spectra match our models extremely well.  Furthermore, we might expect an interferometer misalignment to stretch the wavenumber grid, rather than shifting it.  Finally, Fig. \ref{cia} reveals a spectral offset between the absorption peaks between the \citet{85borysow} and the \citet{07orton} CIA data, indicating that the required spectral shift would be model dependent.   The undulation in the residual could therefore be due to uncertainties in the collision-induced absorption data that have yet to be resolved.

%The wavelength grid used for both Voyagers starts at 150.175 cm$^{-1}$ and has a sampling interval of 1.39051 cm$^{-1}$.  

\subsection{Radiometric calibration of FORCAST}
\label{sec_cal}
\begin{figure*}
\begin{centering}
\centerline{\includegraphics[angle=0,scale=.80]{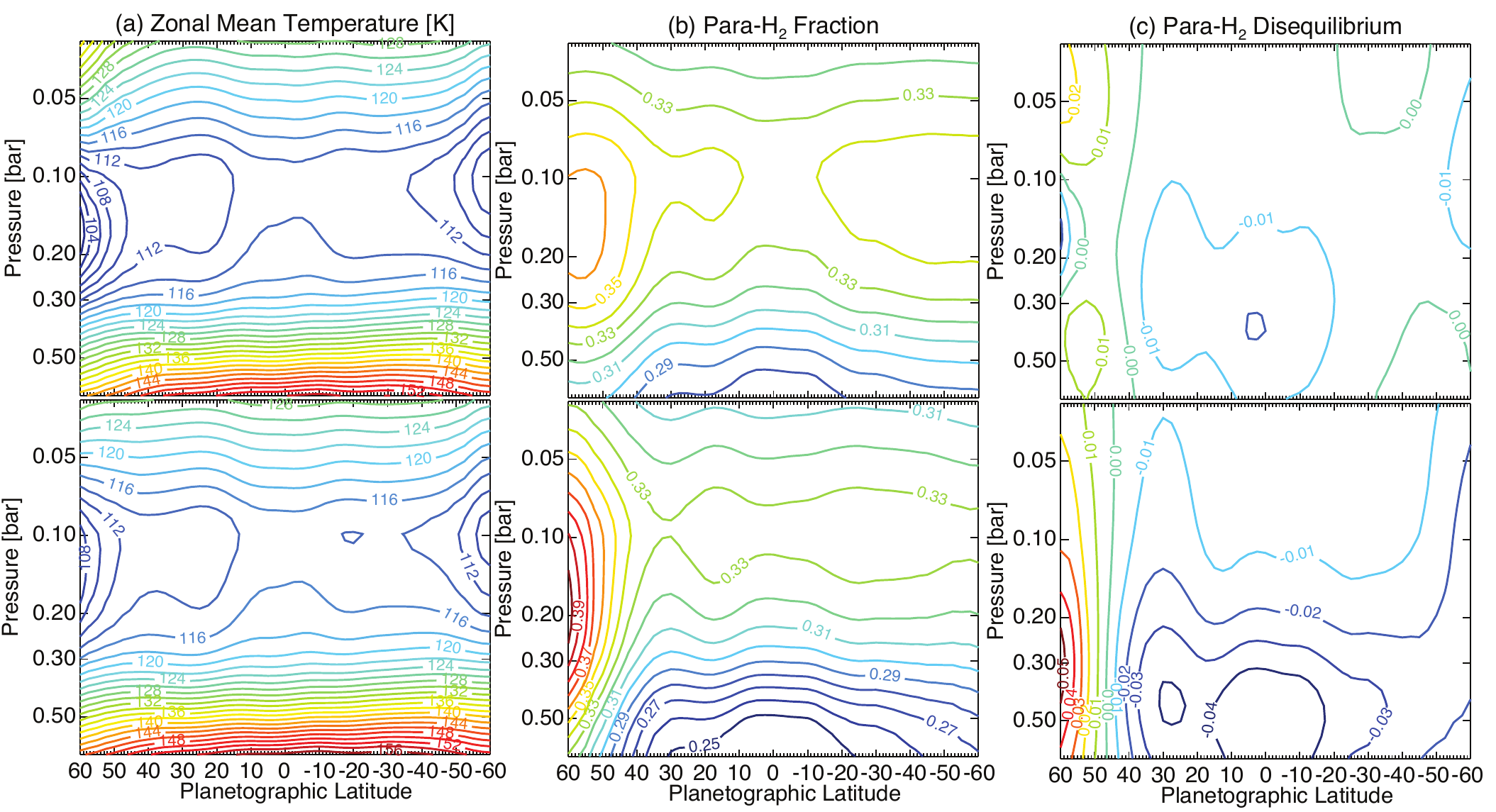}}
\caption{Effect of uniformly scaling the SOFIA/FORCAST flux downwards by 10\% (top row) compared to the original calibration of the data (bottom row).  Temperature, para-H$_2$ and the disqequilibrium are plotted with exactly the same contour spacing as Fig. \ref{comp_all}. }
\label{sofia_scale}
\end{centering}
\end{figure*}

Fig. \ref{comp_all}a demonstrates that the temperatures derived from SOFIA/FORCAST spectra were systematically warmer, by $\sim2$ K at the tropopause and $8-10$ K at 700 mbar, than those obtained by Voyager/IRIS 35 years earlier.  Rather than invoking global changes to Jupiter's temperatures, we assume that this is caused by systematic offsets in the radiometric calibration.  A 10\% reduction of the SOFIA/FORCAST flux makes SOFIA and Voyager approximately consistent with one another (Fig. \ref{sofia_scale}), so we investigate the impact of such a change on the retrieved quantities, repeating the retrievals with scale factor of 0.9, 1.0 and 1.1.  These adjustments are within the 2-12\% uncertainty range associated with the FORCAST absolute calibration (Table \ref{tab:images}).  This results in equatorial 110-mbar temperatures ranging from 114-118 K and equatorial 330-mbar temperatures ranging from 120-123 K.  These are within the $\pm2$ K uncertainty range on the temperature retrieval.  

The absolute para-H$_2$ fraction is larger (and closer to equilibrium) for the $0.9\times$ scale factor (Fig. \ref{sofia_scale}b) by 0.02-0.03 compared to the $1.0\times$ scale factor, meaning that the downward scaling of the SOFIA/FORCAST flux is more consistent with equilibrium conditions (Fig. \ref{sofia_scale}c).   The formal retrieval uncertainties on the FORCAST para-H$_2$ inversions are latitude-dependent, but at mid-latitudes they range from a minimum of 0.014 at 300 mbar to 0.022 at 100 and 750 mbar, excluding systematic uncertainties due to aerosols, helium, radiometric offsets, etc.  At first glance, the small uncertainties might appear to be at odds with the spectral forward models in Fig. \ref{specfits}, where we showed the spectral effects of a $\pm0.05$ offset in the para-H$_2$ fraction.  The resulting models appeared to bracket the $1\sigma$ measurement uncertainties.  However, the para-H$_2$ fraction affects the full 17-37 $\mu$m range, and the $\pm0.05$ offset curves were systematically above or below the best-fitting models, having a substantial effect on the goodness-of-fit.  This means that our formal para-H$_2$ uncertainty is less than 0.05, and closer to the 0.01-0.02 range quoted here.   Nevertheless, this implies that the measured disequilibrium in Fig. \ref{sofia_scale}c is closer to zero within the uncertainties for the $0.9\times$ scale factor, albeit with the same equator-to-pole increase in $f_p$.  Absolute offsets of $\sim4$ K remain present near 700 mbar, but these are not considered robust as they are primarily driven by noisy data from the long-wave FORCAST grism in Fig. \ref{specfits}.

Furthermore, the absolute values of the temperature and para-H$_2$ fraction are sensitive to the degree of smoothing applied to the vertical profiles during the inversion process.  Relax this smoothing, and the temperature and $f_p$ excursions from the \textit{a priori} increase in size until oscillations in the vertical profile become unrealistic and non-physical.  The final result is therefore a compromise between vertical smoothing and the quality of the fit to the data, using \textit{a priori} uncertainties that are based on previous studies - latitudinal temperature excursions of $<5$ K at 250 mbar \citep[based on Cassini observations at higher spatial resolution, e.g.,][]{06simon} and latitudinal para-H$_2$ excursions of $<0.05$ at 250 mbar \citep{98conrath}.  Unfortunately, however, the constraint on the retrieval imposed by the data depends on the signal-to-noise ratio, which is different for Voyager 1, Voyager 2 and the SOFIA/FORCAST data.  The results in Fig. \ref{comp_all} are our best attempt at balancing these competing influences on the retrieval, but a more robust comparison would require the same instrumentation for each epoch under comparison.

%%%%%%%%%%%%%%%%%%%%%%%%%%%%%%%%%%%%%%%%%%%%%%
%%%%%%%%%%%%%%%%%%%%%%%%%%%%%%%%%%%%%%%%%%%%%%
%%%%%%%%%%%%%%%%%%%%%%%%%%%%%%%%%%%%%%%%%%%%%%
\section{Discussion}
\label{discuss}

\begin{figure*}
\begin{centering}
\centerline{\includegraphics[angle=0,scale=.80]{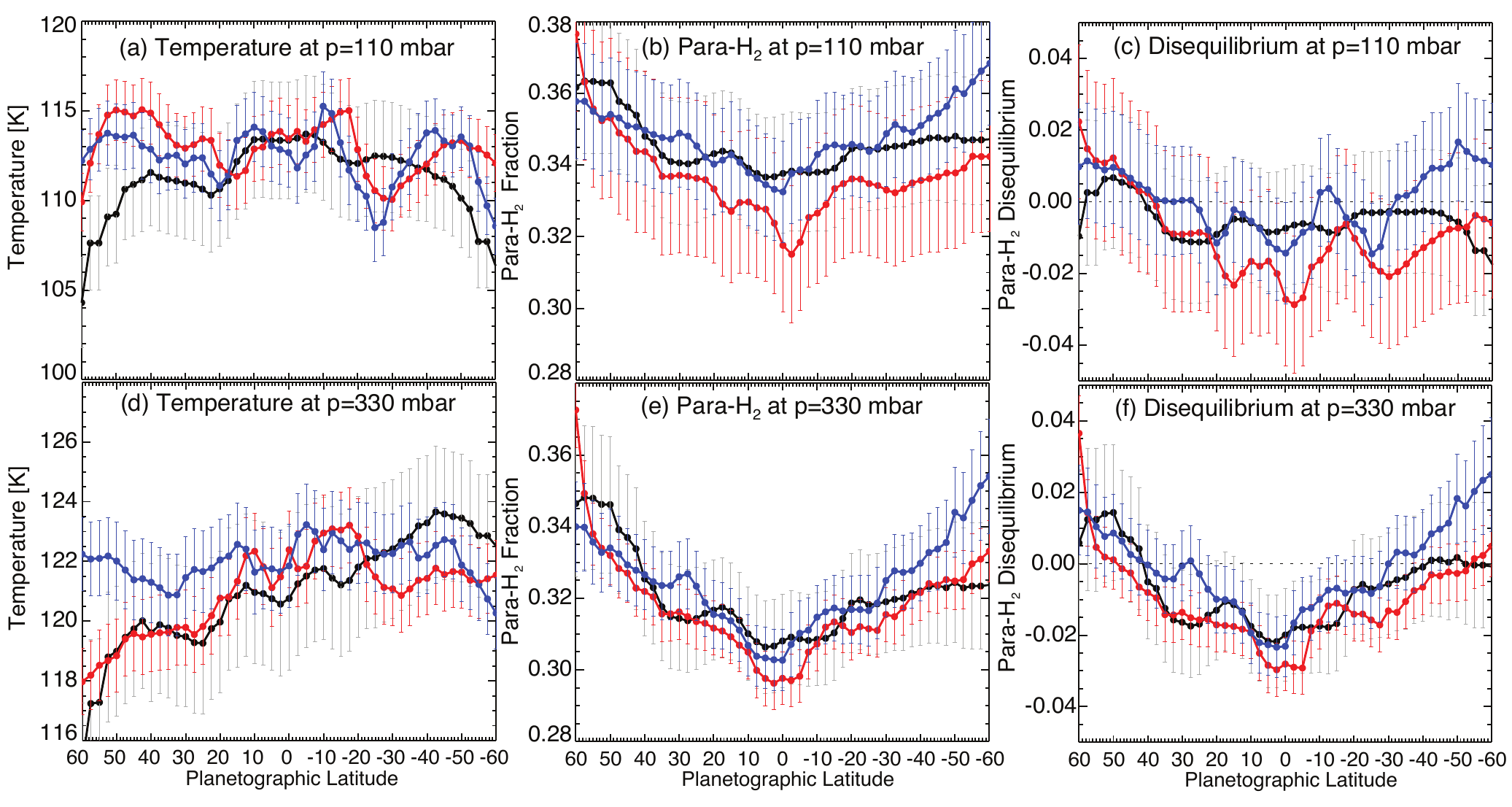}}
\caption{Zonal-mean cross-sections of temperatures, para-H$_2$ fraction and disequilibrium at 110 and 330 mbar.  Black lines (with grey error bars) are from SOFIA/FORCAST with the $0.9\times$ scaling, red lines are from the Voyager-1 north-south maps, and the blue lines are from the Voyager-2 north-south maps.}
\label{comp_final}
\end{centering}
\end{figure*}

Interpretation of spatial variations in Jupiter's far-infrared spectrum remains a substantial challenge almost four decades after the Voyager/IRIS observations, despite the improvements in Earth-based observational capabilities at other wavelengths.  The SOFIA/FORCAST observations represent an attempt to plug this capability gap, but - like Voyager/IRIS - can only resolve the largest scales of Jupiter's banded structure and the Great Red Spot.  Section \ref{models} showed how absolute temperature and para-H$_2$ estimates are adversely affected by uncertainties in radiometric calibration, the chosen source of spectral model for the collision-induced opacity, the degree of smoothing imposed on the spectral inversion, and (to a lesser extent) the influence of aerosol opacity and spectral windowing.  In addition, the helium mole fraction remains poorly constrained and could serve to systematically scale the 17-37 $\mu$m brightness temperatures up and down, which would have the effect of modifying the magnitude of the para-H$_2$ disequilibrium \citep{81gautier}.  The problem is made even more complex by using different instruments with different sensitivities, even between the two Voyager flybys.  Following Section \ref{sec_cal}, we estimate that temperatures and para-H$_2$ fractions are no more accurate than $\pm2$ K and $\pm0.02$ in the 100-700 mbar range, respectively \citep[using the Galileo-derived He abundance and CIA of][]{85borysow}.

Nevertheless, Fig. \ref{comp_all} and Fig. \ref{sofia_scale} do show consistent features for the three epochs studied (March 1979, June 1979 and May 2014), irrespective of the inversion assumptions.  To make this clearer, Fig. \ref{comp_final} extracts zonal-mean cross-sections of temperature, para-H$_2$ and the degree of disequilibrium at 110-mbar (near the tropopause, and at the limit of the vertical sensitivity in Fig. \ref{jacobians}a) and 330 mbar (near the peak of the Jacobians) from SOFIA and the two Voyager north-south sequences.  These all show: an equatorial region ($\pm0-15^\circ$) that is warmer than temperate mid-latitudes ($\pm20-40^\circ$); a temperature drop poleward of $\pm50^\circ$; sub-equilibrium para-H$_2$ in the equatorial region and an equator-to-pole increase in $f_p$ to bring it closer to equilibrium at the highest latitudes.  We discuss each of these findings below.

%%%%%%%%%%%%%%%%%%%%%%%%%%%%%%%%%%%%%%%%%%%%%%
\subsection{Tropospheric Temperatures}
Jupiter's temperature field has been studied at higher spatial resolutions in the mid-infrared from Cassini and ground-based observatories \citep[e.g.,][]{04flasar_jup, 06simon, 11fletcher_fade}.  These revealed equatorial temperatures within $7^\circ$ of the equator that were cooler than the neighbouring northern ($7-17^\circ$N) and southern ($7-20^\circ$S) equatorial belts (NEB, SEB).  In the majority of the SOFIA and Voyager temperature cross-sections in Fig. \ref{comp_all}, these appear to be blended together to form a warm band throughout the $\pm15^\circ$ region.  Fig. 6 of \citet{98conrath} and Fig. 2 of \citet{06simon} showed the same effect.  The raw brightness temperatures in Figs. \ref{spxcontour}-\ref{Tbcompare}, as well as the contextual imaging in Fig. \ref{images}, suggests that there are maxima near $\pm10^\circ$ latitude.  However, these belts have such a limited contrast with the cooler equatorial zone that they have very little influence on the spectral inversions.  The cool equator is certainly present in the FORCAST and IRIS data, but has a limited visibility on the $2.5^\circ$ latitude grid (with $5^\circ$ bin widths) used to analyse the data.  In conclusion, there is no inconsistency with the mid-IR retrievals, but we are left requiring higher spatial resolutions for future far-IR studies of Jupiter's tropics.

The equatorial temperatures observed by FORCAST in 2014 ($\sim115$ K at 200 mbar at the equator) are closest to those measured by Voyager 1, but the differences are within our 1-2 K uncertainty range.  Jupiter's equatorial temperatures are known to vary with time \citep[by 2-4 K at the $\sim250$ mbar level,][]{94orton, 06simon}, possibly as a result of sporadic equatorial brightening events that bring fresh updrafts of material to the tropopause.  \citet{86gierasch} found equatorial differences between Voyager 1 and 2 that are confirmed by Fig. \ref{comp_all}c-d, with the latter being $\sim2$ K cooler at 150 mbar.  This was consistent with the equatorial variability in the ground-based record of \citet{94orton}, which showed the equatorial temperatures cooling by $\sim3$ K between 1978 and 1983.  

At temperate latitudes we do not resolve the belt/zone structure in either the images (Fig. \ref{images}) or the spectra.  The result is a broad cold band at $20-30^\circ$ in both hemispheres, blending together the multiple belts and zones in Jupiter's temperate domain.  During the Voyager encounters, the southern cold band was colder and wider in latitudinal extent than the northern cold band, with a southerly warm belt near $40-50^\circ$ marking the transition into the polar environment.  This asymmetry is reversed in the FORCAST data, with cooler temperatures at northern mid-latitudes than in the south, which can be seen in the raw data (Fig. \ref{spxcontour}a) as an asymmetry between the northern and southern hemispheres.  The changes are at the 2-3 K level, and are consistent with the levels of temperature variability shown in the long-term records of \citet{94orton} and \citet{06simon}.  Finally, we note that all of our inversions (IRIS and FORCAST) suggest a transition to cold polar vortices poleward of $\pm50^\circ$ ($100<p<300$ mbar), consistent with previous studies.  In summary, we find that although FORCAST can offer constraints on Jupiter's tropospheric temperatures, the spatial resolution of SOFIA (and the Voyager observations at far-IR wavelengths) blends together much of the finer belt/zone structures required for dynamical studies.

%%%%%%%%%%%%%%%%%%%%%%%%%%%%%%%%%%%%%%%%%%%%%%
\subsection{Para-Hydrogen}
Despite the low spatial resolution, the spectral coverage of FORCAST and IRIS offer the capability to map Jupiter's para-H$_2$ distribution using the broad S(0) and S(1) lines.  This is a challenging measurement given the different sensitivities of the two instruments, but inversions in Figs. \ref{comp_all} and \ref{comp_final}b show consistent trends with latitude - a sub-equilibrium minimum in $f_p$ within $\sim15^\circ$ of the equator that extends vertically throughout the range of sensitivity (approximately 200-500 mbar), and a transition to equilibrium (and possibly super-equilibrium) conditions at high latitudes.  The para-H$_2$ distribution appears to be asymmetric between the northern and southern hemispheres - FORCAST and Voyager 1 (both observing before northern autumnal equinox) suggest a more rapid increase of $f_p$ with latitude in the northern hemisphere than the southern hemisphere, resulting in super-equilibrium conditions poleward of $50^\circ$N.  Conversely, Voyager 2 observations (taken at northern autumnal equinox) suggest a stronger increase in $f_p$ towards the southern pole.  The latitudinal gradients of the raw brightness temperatures in Fig. \ref{spxcontour} support these findings, with a stronger northern gradient in Voyager 1 observations and a stronger southern gradient in Voyager 2 observations.  The para-H$_2$ distributions do show finer-scale variations with latitude, with some evidence for additional $f_p$ minima in the mid-latitude regions associated with the cool temperate bands described earlier. However, given the retrieval uncertainties of 0.01-0.02 in the $\sim250$-mbar region, these contrasts are not considered to be significant.  

Qualitatively, our Voyager 1 and FORCAST results are consistent with those of \citet{98conrath}, who also showed an asymmetry in $f_p$ and potential super-equilibrium conditions at Jupiter's high northern latitudes (their Fig. 9).  The region of high-$f_p$ air is therefore a repeatable feature of Jupiter's high latitudes, but apparently not a permanent one, with observations by Voyager 2 showing no evidence for this feature in the north.  However, we caution the reader that the fits to the Voyager 2 spectra showed a larger undulation in the residual than Voyager 1 (Section \ref{models}), suggestive of a discrepancy in the wavelength calibration that could influence the retrieved $f_p$ structure.  The time evolution of this equator-to-pole gradient in para-H$_2$ will only be truly revealed by observing Jupiter's poles over longer time spans with the same instrument.

\subsubsection{Balancing dynamics and chemical equilibration}

The vertical distribution of para-H$_2$ is approximately governed by a balance between the strength of vertical mixing and the rate of chemical equilibration between the two spin isomers \citep{86gierasch, 98conrath}.   Unfortunately, neither the timescale for dynamical motions ($\tau_d$) nor the hydrogen equilibration time ($\tau_p$) are well known, and both are expected to vary with latitude.  Hydrogen conversion can occur in the gas phase due to (i) hydrogen exchange between molecular and atomic hydrogen or (ii) paramagnetic conversion between two H$_2$ molecules, but these are deemed less efficient (i.e., longer $\tau_p$) than (iii) paramagnetic conversion on the surfaces of aerosols \citep{82massie, 84conrath, 03fouchet}.  The efficiency of conversion depends on many factors - the surface area of the aerosols; UV irradiation of the aerosols to generate active surface sites; their optical depth; and the mixing of the air within the cloud layers.  If the para-H$_2$ distribution solely reflected the efficiency of this paramagnetic conversion and was not influenced by vertical mixing, then we should find that the cloudiest regions (i.e., the equator and the northern and southern tropical zones) would have values closest to equilibrium, with more extreme disequilibrium over the cloud-free belts.  \citet{92carlson}, in their analysis of a small number of localised regions in the IRIS dataset, suggested that such a correlation did exist.  However, we find that Jupiter's equator is the site of the largest sub-equilibrium conditions, counter to the purely chemical explanation.  

This led \citet{98conrath} to provide a quantitative assessment of the timescales governing the distribution of para-H$_2$ taking dynamics into account, showing that that the dynamical timescale must be smaller ($\sim70$ years) than the timescale required for the equilibration between the ortho- and para-H$_2$ ($\sim110$ years) in Jupiter's upper troposphere.   Parcels of low-$f_p$ air could be advected upwards into colder atmospheric layers at a rate faster than the conversion to ortho-H$_2$.  However, \citep{03fouchet} scaled their stratospheric para-H$_2$ results to the 200-mbar level and suggested lower limits on $\tau_p$ of 15-22 years assuming paramagnetic conversion on aerosols, or 30-50 years assuming paramagnetic conversion in the gas phase (they did not favour hydrogen exchange as a dominant mechanism in Jupiter's troposphere, as this resulted in extremely long equilibration timescales).  These lower limits raise the possibility that $\tau_p$ could be smaller than $\tau_d$ on Jupiter \citep[as found on the other giant planets by][]{98conrath}, suggesting that chemical conversion could indeed dominate over dynamics.  Further quantitative progress in separating the dynamical and chemical influences on $f_p(p)$ cannot be made without better constraints on the hydrogen conversion mechanism.

\subsubsection{High-latitude para-H$_2$}

The polar tropopause is colder than elsewhere on the planet (Fig. \ref{comp_all}), meaning that the equilibrium para-H$_2$ is largest at the poles.  However, we consistently retrieve $f_p$ in excess of this cold equilibrium in both the SOFIA and Voyager datasets.  This matches the findings of \citet{98conrath}, who suggested that these high-latitude $f_p$ maxima were due to air descending below the tropopause, enriching the 200-500 mbar levels in high-$f_p$ air from the colder layers above.  Such a circulation pattern, with air rising at low latitudes and descending over the poles, is consistent with that modelled by \citet{90conrath}.  However, given the uncertainties on the high-latitude super-equilibrium shown in Fig. \ref{comp_final}, we must question the validity of this conclusion.

The inversions in Fig. \ref{comp_all} placed the largest $f_p$ in the 50-100 mbar region above the tropopause, which makes little sense as the equilibrium $f_p$ should decline with altitude in the lower stratosphere as the temperature rises.  These lower-stratospheric values are likely to be untrustworthy, given the poor vertical sensitivity of the far-IR inversions to the tropopause regions (Fig. \ref{jacobians}), and the high emission angles that were used to sample the $50-60^\circ$ latitude regions.  These could lead to systematic errors in the para-H$_2$ retrieval.  However, the stratospheric investigation of \citet{03fouchet} suggested the possibility that Jupiter's stratospheric $f_p$ matches the tropopause value (at least in the $\pm30^\circ$ latitude range averaged in their ISO observations), i.e., Jupiter's $f_p$ remains larger than equilibrium in the stratosphere without any conversion back to ortho-H$_2$. The vertical $T(p)$ is poorly known at high latitudes and $p<100$ mbar, particularly within the cold polar vortex, so maybe a region of high-$f_p$ air could exist at lower pressures. The hypothesis that subsidence generates regions of high-$f_p$ air at high latitudes on Jupiter cannot be rejected by these data.

However, Fig. \ref{comp_final} shows that the strongest conclusion that can be drawn is that $f_p$ is closest to equilibrium at Jupiter's high latitudes, so large-scale subsidence need not be invoked. Instead, if we assume the efficiency of hydrogen conversion to be latitudinally uniform, then the equator-to-pole gradient may simply reflect weaker tropospheric upwelling at high latitudes to replenish equilibrated air near the tropopause.  Indeed, parameterisations of Jupiter's eddy diffusivity by \citet{15wang} demonstrate a strong decrease in the expected strength of vertical mixing from the equator to pole due to the planet's rotation.  We investigate correlations with other tracers of tropospheric mixing in the next section.

\subsection{Comparing para-H$_2$ to tropospheric gases and aerosols}

\begin{figure}
\begin{centering}
\centerline{\includegraphics[angle=0,scale=.90]{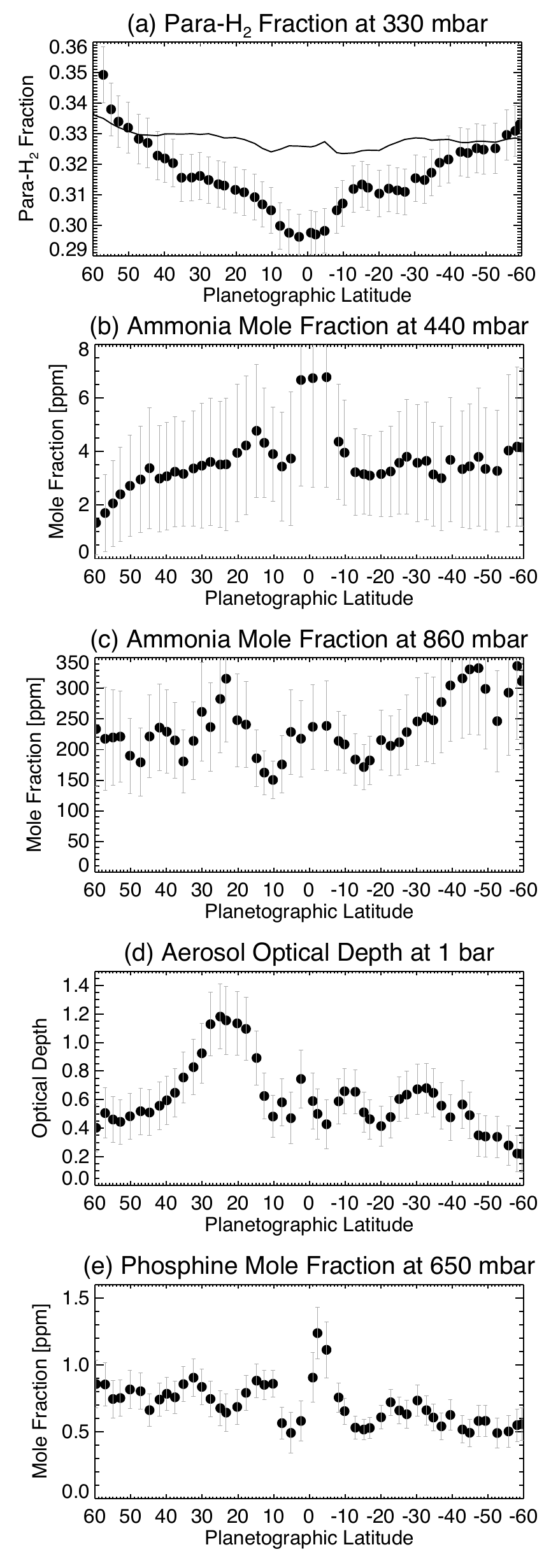}}
\caption{Comparison of para-H$_2$ with tropospheric properties derived from Voyager-1/IRIS spectra.  The solid line in panel (a) shows the equilibrium para-H$_2$ fraction based on the retrieved temperatures, showing that the majority of Jupiter's atmosphere is sub-equilibrium.  Panels b-e show the ammonia, phosphine and aerosol distributions derived from the full 300-1350 cm$^{-1}$ IRIS spectrum (all previous figures have used only the 300-570 cm$^{-1}$ region). }
\label{gases}
\end{centering}
\end{figure}

To better explore the correlation between para-H$_2$ and Jupiter's other tropospheric variables, we replicate the analysis of \citet{86gierasch} by extending the Voyager-1/IRIS spectral retrieval to cover the full 300-1350 cm$^{-1}$ (7.4-33.3 $\mu$m) range (all the aforementioned results considered 300-600 cm$^{-1}$ only).  This allows us to study Jupiter's NH$_3$, PH$_3$ and aerosol opacity in the mid-IR (see Jacobians in Fig. \ref{jacobians}c-d) at the same epoch and spatial resolution as the temperatures and para-H$_2$ discussed above.  Jupiter's $T(p)$ and para-H$_2$ distribution were retrieved simultaneously with a parameterised NH$_3$ distribution (a variable deep abundance for $p>800$ mbar, declining at higher altitudes according to a variable fractional scale height), and a scale factor for the PH$_3$ distribution of \citet{09fletcher_ph3}. In addition, we scaled the optical depth of an 800-mbar compact cloud layer consisting of NH$_3$ ice particles \citep[refractive indices based on][]{84martonchik} with a log-normal distribution of radius $10\pm5$ $\mu$m.  Aerosol cross sections and single scattering albedos were calculated assuming Mie scattering. The sources of spectral line data are identical to those used for Cassini/CIRS analyses in \citet{09fletcher_ph3}, and were converted to $k$-distributions for IRIS modelling using a Hamming line shape of full-width-at-half-maximum of 4.3 cm$^{-1}$.  The distribution of para-H$_2$ at 330 mbar is compared to ammonia, phosphine and aerosols from Voyager-1/IRIS in Fig. \ref{gases}.  

Firstly, we note that the zonal-mean NH$_3$ distributions in Fig. \ref{gases}b-c largely reproduce those found by \citet{86gierasch} using the same dataset, and have mole fractions comparable to those reported from Cassini \citep{06achterberg}.  The para-H$_2$ distribution also resembles that shown in Fig. 5 of \citet{86gierasch}.  Phosphine in Fig. \ref{gases}e is found to be enhanced near the equator compared to the neighbouring belts, as confirmed later by Cassini \citep{04irwin, 09fletcher_ph3}.  Both PH$_3$ and NH$_3$ show low-latitude enhancements in the regions of sub-equilibrium para-H$_2$, suggesting that they are being enriched by the same upwelling motions that are responsible for the low-$f_p$ air at the equator.

The aerosol optical depth in Fig. \ref{gases}d was derived using the 600-1350 cm$^{-1}$ (7.4-16.7 $\mu$m) region alone to ensure that it was independent of any degeneracies between aerosols and para-H$_2$ in the 300-600 cm$^{-1}$ region.  We recover the latitudinal variation of optical depth identified by \citet{86gierasch} (their Figure 4), including the high opacity in the $20-30^\circ$N band.  Intriguingly, this does not closely resemble the cloud opacity distributions derived later by Cassini \citep[e.g.,][]{05matcheva, 16fletcher_texes}, which showed distinct maxima at the equator and the NTrZ and STrZ (northern and southern tropical zones, respectively) and strong minima associated with the NEB and SEB (northern and southern equatorial belts).  The absence of the equatorial aerosol enhancement could be related to spatial resolution - the cold equatorial troposphere is also not well resolved by the IRIS data in Fig. \ref{comp_all}.  

However, Fig. \ref{gases} makes it clear that para-H$_2$ is not well correlated with Jupiter's tropospheric aerosols, ammonia and phosphine distributions.  None of these species reflect a decline in the strength of vertical mixing with latitude \citep[e.g.,][]{15wang}, which again raises the possibility that chemical equilibration, rather than dynamic mixing, governs the equator-to-pole para-H$_2$ gradient. Our experiments do not yield the correlation between para-H$_2$ and cloud opacity favoured by \citet{92carlson}, but given the limited sensitivity of thermal-IR spectra to aerosols, we cannot completely reject the aerosol-catalysis hypothesis.

Despite the lack of correlation with the tropospheric aerosols in the main cloud deck (Fig. \ref{gases}d), interconversion of para- and ortho-H$_2$ on the surfaces of aerosols remains an intriguing possibility that has been favoured by previous studies \citep{92carlson, 03fouchet}. The character of Jupiter's stratospheric aerosols is known to change at the highest latitudes.  Inversions of Cassini/ISS visible-light observations by \citet{13zhang_aer} show that the number density and optical depth of stratospheric aerosols increases by orders of magnitude from the equator to the poles (with a peak in the 20-50 mbar region), and that this haze extends down to the tropopause in the $\pm60^\circ$ latitude region.  The increased availability of these upper tropospheric and stratospheric hazes at Jupiter's high latitudes could increase the efficiency of aerosol catalysis there, explaining why para-H$_2$ is closer to equilibrium at high latitudes compared to low latitudes.   Furthermore, there is a subtle asymmetry in the stratospheric haze optical thickness \citep[Fig. 12 of][]{13zhang_aer}, with more aerosol loading in the northern hemisphere than the south, potentially as a result of the auroral generation mechanism.  If the high-latitude $f_p$ enrichment is indeed linked to the hazes, and not to the deeper tropospheric clouds, then this could explain the observed asymmetry.  Mid- and far-infrared spectral inversions are insensitive to these haze particles because of their small radii and low optical depths, so they could not be derived from the IRIS and FORCAST data.

At Jupiter's high latitudes, we propose that $f_p$ is close to equilibrium due to the efficiency of equilibration on the upper tropospheric/stratospheric aerosols compared to low-latitudes, potentially with super-equilibrium conditions created by subsidence over the poles (provided a region of high-$f_p$ air resides above the altitude range of far-IR sensitivity).  Conversely, the low-latitude sub-equilibrium $f_p$ is due to strong vertical mixing in the tropics, as well as inefficient hydrogen equilibration due to a clearer upper troposphere.  In this regard, the large-scale equator-to-pole gradient is related to the availability of aerosols to catalyse the equilibration, whereas the local-scale features (the strong perturbations at the equator and high latitudes) are the result of vertical advection and mixing.  This upwelling might not be uniform over the equatorial zone, but confined to temporally-variable plume activity, similar to the low-$f_p$ air that was produced by Saturn's northern springtime storm \citep{14achterberg}.  Higher resolution observations of $f_p$ at all latitudes will be required to better correlate para-H$_2$ with dynamic phenomena.

Finally, Saturn also exhibits sub-equilibrium conditions at the equator, but with high-latitude $f_p$ that changes with time \citep{16fletcher}.  During Cassini's observations of Saturn, a region of high-$f_p$ air has been steadily developing in Saturn's northern hemisphere (poleward of $\sim50^\circ$N) as Saturn passed northern spring equinox and approached summer solstice.  \citet{16fletcher} suggested that this was due to tropospheric subsidence in the springtime hemisphere and that it could be related to the growth of springtime aerosols.  Jupiter's own $f_p$ is unlikely to vary seasonally due to the low obliquity, which would suggest that the observed asymmetry in the FORCAST and Voyager-1 data could be relatively stable.  However, we have no explanation for why the high-$f_p$ airmass is absent from the Voyager-2 observations of the northern hemisphere.

%%%%%%%%%%%%%%%%%%%%%%%%%%%%%%%%%%%%%%%%%%%%%%
%%%%%%%%%%%%%%%%%%%%%%%%%%%%%%%%%%%%%%%%%%%%%%
%%%%%%%%%%%%%%%%%%%%%%%%%%%%%%%%%%%%%%%%%%%%%%
\section{Conclusions}
\label{conclusion}

Observations of Jupiter's far-infrared spectrum from the FORCAST instrument on the SOFIA aircraft in May 2014 have provided spectroscopic maps of Jupiter with a 2-4" spatial resolution.  The spectra cover the hydrogen-helium collision-induced absorption in the 17-37 $\mu$m region that is largely inaccessible to ground-based observatories.  These spectra have been inverted to determine Jupiter's temperature and para-H$_2$ distribution as a function of latitude and pressure in the 70-700 mbar region.  Despite the low spatial resolution, this is the first time that such a measurement has been possible since the Voyager flybys of Jupiter in 1979, and permit a quantitative comparison between the two epochs.  

Jupiter's disc-integrated brightness varies from $122.8\pm3.7$ K at 31 $\mu$m to $136.2\pm4.7$ K at 37.1 $\mu$m.  Tropospheric temperatures observed by FORCAST agree with those found by Voyager, and show the following characteristics:  (i) Jupiter's cold equatorial zone has limited influence on the far-IR spectra and shows low contrast with the neighbouring warm equatorial belts; (ii) the temperate banded structure is blended together to form broad, cool zones between $20-40^\circ$ which have varied in latitudinal size between the Voyager and SOFIA datasets; (iii) longitudinal undulations of tropospheric brightness on the NEB are well-correlated between the M (5.4 $\mu$m) and N-band (8-11 $\mu$m) observations, suggestive of wave activity affecting cloud opacity, tropospheric temperatures, and ammonia humidity throughout a broad region of the atmosphere between the tropopause and the cloud decks; (iv) both the Voyager and SOFIA inversions suggest the presence of cold vortices poleward of $\pm50^\circ$ latitude and $100<p<300$ mbar; and (v) stratospheric heating associated with the northern auroral oval is detectable by SOFIA as increased brightness near $180^\circ$ in the 7.7- and 11.1-$\mu$m filters sensing stratospheric methane and ethane, respectively.  

Many of these features are better resolved at mid-infrared wavelengths with larger observatories, although the key strength of SOFIA is the ability to derive the para-H$_2$ distribution from the relative strength of the S(0) and S(1) absorption features.  The SOFIA results are comparable with previous Voyager findings that show an equator-to-pole increase in the para-H$_2$ fraction, with low-$f_p$ and sub-equilibrium conditions at the equator and high-$f_p$ and weak super-equilibrium conditions at $\pm60^\circ$ latitude.  We confirm the presence of a region of high-$f_p$ air at high northern latitudes that appears asymmetric when compared to the south, which was previously seen by Voyager 1 \citep{86gierasch, 98conrath}.  This region is therefore a repeatable feature, although it does not appear to be permanent (it was not evident in the noisier Voyager 2 data), and the magnitude of the super-equilibrium depends on our retrieval assumptions.  We note that the choice of collision-induced H$_2$-He opacity influences our quantitative results, and that the use of recent opacity tables \citep{07orton,12richard} worsens the fits to FORCAST and IRIS spectra when compared to the original (erroneous) opacity results from \citet{85borysow}.  We suggest that this is due to missing (H$_2$)$_2$ dimer absorption in these tables, which increases the absorption coefficient near the S(0) and S(1) lines to values closer to those of \citet{85borysow}.

Inversions of Voyager-1 mid-infrared spectra revealed that the latitudinal distributions of tropospheric ammonia, phosphine and cloud opacity at $\sim800$ mbar are not correlated with the observed distribution of para-H$_2$ in the SOFIA and Voyager observations.  Nevertheless, we note a similarity between the para-H$_2$ distribution and that of small-radii aerosols comprising the upper tropospheric and stratospheric hazes \citep{13zhang_aer}.  These hazes are also asymmetric, with a greater aerosol mass loading in the north polar regions where we find the coldest tropopause temperatures and the highest para-H$_2$ fraction.  We therefore propose that the para-H$_2$ equator-to-pole gradient is governed primarily by the efficiency of paramagnetic conversion catalysed by aerosols in Jupiter's hazes (rather than in the cloud decks sensed in the mid- and far-infrared); with secondary perturbations from localised circulations.  In this scenario, the sub-equilibrium conditions at low latitudes are due to the clear-atmosphere conditions and inefficient equilibration, coupled with strong upwelling enriching low-$f_p$ air at the equator.  Conversely, high-latitude air is closer to equilibrium due to efficient hydrogen conversion and weaker upwelling, and may be locally enhanced in the north by subsidence beneath the cold polar vortex.

Observations of Jupiter at higher spatial resolutions could start to distinguish the competing dynamical and chemical influences on the para-H$_2$ distribution, particularly if $f_p$ is found to be substantially sub-equilibrium in localised plumes at the equator and other latitudes, such as those found on Saturn \citep{14achterberg}.  Furthermore, the strength of the para-H$_2$ asymmetry and the north polar anomaly could be tracked with a time series of measurements at various seasons during Jupiter's 11.8-year orbit.  Unfortunately Jupiter's H$_2$ continuum is beyond the saturation limits for the James Webb Space Telescope \citep[which could have reached the S(0) line near 350 cm$^{-1}$ (28.5 $\mu$m),][]{16norwood}, although this facility should certainly be used to conduct similar studies of the para-H$_2$ distributions on Uranus and Neptune.

%%%%%%%%%%%%%%%%%%%%%%%%%%%%%%%%
%%%%%%%%%%%%%%%%%%%%%%%%%%%%%%%%
%%%%%%%%%%%%%%%%%%%%%%%%%%%%%%%%
\section*{Acknowledgments}

This work was based on observations made with the NASA/DLR Stratospheric Observatory for Infrared Astronomy (SOFIA), and was financially supported through SOF0012 to the University of California, Berkeley.  SOFIA is jointly operated by the Universities Space Research Association, Inc. (USRA), under NASA contract NAS2-97001, and the Deutsches SOFIA Institut (DSI) under DLR contract 50 OK 0901 to the University of Stuttgart.  We are grateful for all those involved in the telescope engineering, operations support and the flight crews.  We thank Luke Keller, Matthew Bellardini and Joseph Quinn (Ithaca College) and Joseph Adams for their assistance with the initial planning and calibration of the SOFIA data.  Fletcher was supported by a Royal Society Research Fellowship at the University of Leicester.  The UK authors acknowledge the support of the Science and Technology Facilities Council (STFC). A portion of this work was performed by Orton at the Jet Propulsion Laboratory, California Institute of Technology, under a contract with NASA.  Gehrz received partial support from the United States Air Force.  This research used the ALICE High Performance Computing Facility at the University of Leicester.

\bibliographystyle{elsarticle-harv}
\bibliography{references}

\begin{thebibliography}{71}
\expandafter\ifx\csname natexlab\endcsname\relax\def\natexlab#1{#1}\fi
\providecommand{\url}[1]{\texttt{#1}}
\providecommand{\href}[2]{#2}
\providecommand{\path}[1]{#1}
\providecommand{\DOIprefix}{doi:}
\providecommand{\ArXivprefix}{arXiv:}
\providecommand{\URLprefix}{URL: }
\providecommand{\Pubmedprefix}{pmid:}
\providecommand{\doi}[1]{\href{http://dx.doi.org/#1}{\path{#1}}}
\providecommand{\Pubmed}[1]{\href{pmid:#1}{\path{#1}}}
\providecommand{\bibinfo}[2]{#2}
\ifx\xfnm\relax \def\xfnm[#1]{\unskip,\space#1}\fi
%Type = Article
\bibitem[{{Achterberg} et~al.(2006){Achterberg}, {Conrath} and
  {Gierasch}}]{06achterberg}
\bibinfo{author}{{Achterberg}, R.K.}, \bibinfo{author}{{Conrath}, B.J.},
  \bibinfo{author}{{Gierasch}, P.J.}, \bibinfo{year}{2006}.
\newblock \bibinfo{title}{{Cassini CIRS retrievals of Ammonia in Jupiter's
  Upper Troposphere}}.
\newblock \bibinfo{journal}{Icarus} \bibinfo{volume}{182},
  \bibinfo{pages}{169--180}.
\newblock \DOIprefix\doi{10.1016/j.icarus.2005.12.020}.
%Type = Article
\bibitem[{{Achterberg} et~al.(2014){Achterberg}, {Gierasch}, {Conrath},
  {Fletcher}, {Hesman}, {Bjoraker} and {Flasar}}]{14achterberg}
\bibinfo{author}{{Achterberg}, R.K.}, \bibinfo{author}{{Gierasch}, P.J.},
  \bibinfo{author}{{Conrath}, B.J.}, \bibinfo{author}{{Fletcher}, L.N.},
  \bibinfo{author}{{Hesman}, B.E.}, \bibinfo{author}{{Bjoraker}, G.L.},
  \bibinfo{author}{{Flasar}, F.M.}, \bibinfo{year}{2014}.
\newblock \bibinfo{title}{{Changes to Saturn's Zonal-mean Tropospheric Thermal
  Structure after the 2010-2011 Northern Hemisphere Storm}}.
\newblock \bibinfo{journal}{ApJ} \bibinfo{volume}{786}, \bibinfo{pages}{92}.
\newblock \DOIprefix\doi{10.1088/0004-637X/786/2/92}.
%Type = Inproceedings
\bibitem[{{Adams} et~al.(2010){Adams}, {Herter}, {Gull}, {Schoenwald},
  {Henderson}, {Keller}, {De Buizer}, {Stacey} and {Nikola}}]{10adams}
\bibinfo{author}{{Adams}, J.D.}, \bibinfo{author}{{Herter}, T.L.},
  \bibinfo{author}{{Gull}, G.E.}, \bibinfo{author}{{Schoenwald}, J.},
  \bibinfo{author}{{Henderson}, C.P.}, \bibinfo{author}{{Keller}, L.D.},
  \bibinfo{author}{{De Buizer}, J.M.}, \bibinfo{author}{{Stacey}, G.J.},
  \bibinfo{author}{{Nikola}, T.}, \bibinfo{year}{2010}.
\newblock \bibinfo{title}{{FORCAST: a first light facility instrument for
  SOFIA}}, in: \bibinfo{booktitle}{Ground-based and Airborne Instrumentation
  for Astronomy III}, p. \bibinfo{pages}{77351U}.
\newblock \DOIprefix\doi{10.1117/12.857049}.
%Type = Article
\bibitem[{{Atreya} et~al.(1999){Atreya}, {Wong}, {Owen}, {Mahaffy}, {Niemann},
  {de Pater}, {Drossart} and {Encrenaz}}]{99atreya}
\bibinfo{author}{{Atreya}, S.K.}, \bibinfo{author}{{Wong}, M.H.},
  \bibinfo{author}{{Owen}, T.C.}, \bibinfo{author}{{Mahaffy}, P.R.},
  \bibinfo{author}{{Niemann}, H.B.}, \bibinfo{author}{{de Pater}, I.},
  \bibinfo{author}{{Drossart}, P.}, \bibinfo{author}{{Encrenaz}, T.},
  \bibinfo{year}{1999}.
\newblock \bibinfo{title}{{A comparison of the atmospheres of Jupiter and
  Saturn: deep atmospheric composition, cloud structure, vertical mixing, and
  origin}}.
\newblock \bibinfo{journal}{Plan. \& Space Sci.} \bibinfo{volume}{47},
  \bibinfo{pages}{1243--1262}.
%Type = Article
\bibitem[{{Birnbaum} et~al.(1996){Birnbaum}, {Borysow} and
  {Orton}}]{96birnbaum}
\bibinfo{author}{{Birnbaum}, G.}, \bibinfo{author}{{Borysow}, A.},
  \bibinfo{author}{{Orton}, G.S.}, \bibinfo{year}{1996}.
\newblock \bibinfo{title}{{Collision-Induced Absorption of H$_2$-H$_2$ and
  H$_2$-He in the Rotational and Fundamental Bands for Planetary
  Applications}}.
\newblock \bibinfo{journal}{Icarus} \bibinfo{volume}{123},
  \bibinfo{pages}{4--22}.
\newblock \DOIprefix\doi{10.1006/icar.1996.0138}.
%Type = Article
\bibitem[{Borysow et~al.(1985)Borysow, Trafton, Frommhold and
  Birnbaum}]{85borysow}
\bibinfo{author}{Borysow, J.}, \bibinfo{author}{Trafton, L.},
  \bibinfo{author}{Frommhold, L.}, \bibinfo{author}{Birnbaum, G.},
  \bibinfo{year}{1985}.
\newblock \bibinfo{title}{Modeling of pressure-induced far-infrared
  absorption-spectra: molecular-hydrogen pairs}.
\newblock \bibinfo{journal}{Astrophys. J.} \bibinfo{volume}{296},
  \bibinfo{pages}{644--654}.
%Type = Inproceedings
\bibitem[{{Burke}(1974)}]{74burke}
\bibinfo{author}{{Burke}, T.E.}, \bibinfo{year}{1974}.
\newblock \bibinfo{title}{{The Mariner Jupiter/Saturn IRIS experiment}}, in:
  \bibinfo{editor}{{Palluconi}, F.D.}, \bibinfo{editor}{{Pettengill}, G.H.}
  (Eds.), \bibinfo{booktitle}{The Saturn's Rings Workshop}.
%Type = Article
\bibitem[{{Caldwell} et~al.(1980){Caldwell}, {Gillett} and
  {Tokunaga}}]{80caldwell}
\bibinfo{author}{{Caldwell}, J.}, \bibinfo{author}{{Gillett}, F.C.},
  \bibinfo{author}{{Tokunaga}, A.T.}, \bibinfo{year}{1980}.
\newblock \bibinfo{title}{{Possible infrared aurorae on Jupiter}}.
\newblock \bibinfo{journal}{Icarus} \bibinfo{volume}{44},
  \bibinfo{pages}{667--675}.
\newblock \DOIprefix\doi{10.1016/0019-1035(80)90135-9}.
%Type = Article
\bibitem[{{Carlson} et~al.(1992){Carlson}, {Lacis} and {Rossow}}]{92carlson}
\bibinfo{author}{{Carlson}, B.E.}, \bibinfo{author}{{Lacis}, A.A.},
  \bibinfo{author}{{Rossow}, W.B.}, \bibinfo{year}{1992}.
\newblock \bibinfo{title}{{Ortho-para-hydrogen equilibration on Jupiter}}.
\newblock \bibinfo{journal}{Astrophys. J.} \bibinfo{volume}{393},
  \bibinfo{pages}{357--372}.
\newblock \DOIprefix\doi{10.1086/171510}.
%Type = Article
\bibitem[{{Conrath} et~al.(1981a){Conrath}, {Flasar}, {Pirraglia}, {Gierasch}
  and {Hunt}}]{81conrath_grs}
\bibinfo{author}{{Conrath}, B.J.}, \bibinfo{author}{{Flasar}, F.M.},
  \bibinfo{author}{{Pirraglia}, J.A.}, \bibinfo{author}{{Gierasch}, P.J.},
  \bibinfo{author}{{Hunt}, G.E.}, \bibinfo{year}{1981}a.
\newblock \bibinfo{title}{{Thermal structure and dynamics of the Jovian
  atmosphere. II - Visible cloud features}}.
\newblock \bibinfo{journal}{Journal of Geophysical Research}
  \bibinfo{volume}{86}, \bibinfo{pages}{8769--8775}.
\newblock \DOIprefix\doi{10.1029/JA086iA10p08769}.
%Type = Inproceedings
\bibitem[{{Conrath} and {Gautier}(1980)}]{80conrath}
\bibinfo{author}{{Conrath}, B.J.}, \bibinfo{author}{{Gautier}, D.},
  \bibinfo{year}{1980}.
\newblock \bibinfo{title}{{Thermal structure of Jupiter's atmosphere obtained
  by inversion of Voyager 1 infrared measurements}}, in:
  \bibinfo{booktitle}{Remote Sensing of Atmospheres and Oceans}, pp.
  \bibinfo{pages}{611--628}.
%Type = Article
\bibitem[{{Conrath} and {Gautier}(2000)}]{00conrath}
\bibinfo{author}{{Conrath}, B.J.}, \bibinfo{author}{{Gautier}, D.},
  \bibinfo{year}{2000}.
\newblock \bibinfo{title}{{Saturn Helium Abundance: A Reanalysis of Voyager
  Measurements}}.
\newblock \bibinfo{journal}{Icarus} \bibinfo{volume}{144},
  \bibinfo{pages}{124--134}.
\newblock \DOIprefix\doi{10.1006/icar.1999.6265}.
%Type = Article
\bibitem[{{Conrath} and {Gierasch}(1984)}]{84conrath}
\bibinfo{author}{{Conrath}, B.J.}, \bibinfo{author}{{Gierasch}, P.J.},
  \bibinfo{year}{1984}.
\newblock \bibinfo{title}{{Global variation of the para hydrogen fraction in
  Jupiter's atmosphere and implications for dynamics on the outer planets}}.
\newblock \bibinfo{journal}{Icarus} \bibinfo{volume}{57},
  \bibinfo{pages}{184--204}.
\newblock \DOIprefix\doi{10.1016/0019-1035(84)90065-4}.
%Type = Article
\bibitem[{{Conrath} et~al.(1990){Conrath}, {Gierasch} and {Leroy}}]{90conrath}
\bibinfo{author}{{Conrath}, B.J.}, \bibinfo{author}{{Gierasch}, P.J.},
  \bibinfo{author}{{Leroy}, S.S.}, \bibinfo{year}{1990}.
\newblock \bibinfo{title}{{Temperature and circulation in the stratosphere of
  the outer planets}}.
\newblock \bibinfo{journal}{Icarus} \bibinfo{volume}{83},
  \bibinfo{pages}{255--281}.
\newblock \DOIprefix\doi{10.1016/0019-1035(90)90068-K}.
%Type = Article
\bibitem[{{Conrath} et~al.(1981b){Conrath}, {Gierasch} and
  {Nath}}]{81conrath_stab}
\bibinfo{author}{{Conrath}, B.J.}, \bibinfo{author}{{Gierasch}, P.J.},
  \bibinfo{author}{{Nath}, N.}, \bibinfo{year}{1981}b.
\newblock \bibinfo{title}{{Stability of zonal flows on Jupiter}}.
\newblock \bibinfo{journal}{Icarus} \bibinfo{volume}{48},
  \bibinfo{pages}{256--282}.
\newblock \DOIprefix\doi{10.1016/0019-1035(81)90108-1}.
%Type = Article
\bibitem[{{Conrath} et~al.(1998){Conrath}, {Gierasch} and
  {Ustinov}}]{98conrath}
\bibinfo{author}{{Conrath}, B.J.}, \bibinfo{author}{{Gierasch}, P.J.},
  \bibinfo{author}{{Ustinov}, E.A.}, \bibinfo{year}{1998}.
\newblock \bibinfo{title}{{Thermal Structure and Para Hydrogen Fraction on the
  Outer Planets from Voyager IRIS Measurements}}.
\newblock \bibinfo{journal}{Icarus} \bibinfo{volume}{135},
  \bibinfo{pages}{501--517}.
\newblock \DOIprefix\doi{10.1006/icar.1998.6000}.
%Type = Article
\bibitem[{{Conrath} and {Pirraglia}(1983)}]{83conrath}
\bibinfo{author}{{Conrath}, B.J.}, \bibinfo{author}{{Pirraglia}, J.A.},
  \bibinfo{year}{1983}.
\newblock \bibinfo{title}{{Thermal structure of Saturn from Voyager infrared
  measurements - Implications for atmospheric dynamics}}.
\newblock \bibinfo{journal}{Icarus} \bibinfo{volume}{53},
  \bibinfo{pages}{286--292}.
\newblock \DOIprefix\doi{10.1016/0019-1035(83)90148-3}.
%Type = Article
\bibitem[{{Flasar} et~al.(1981){Flasar}, {Conrath}, {Pirraglia}, {Clark},
  {French} and {Gierasch}}]{81flasar}
\bibinfo{author}{{Flasar}, F.M.}, \bibinfo{author}{{Conrath}, B.J.},
  \bibinfo{author}{{Pirraglia}, J.}, \bibinfo{author}{{Clark}, P.C.},
  \bibinfo{author}{{French}, R.G.}, \bibinfo{author}{{Gierasch}, P.J.},
  \bibinfo{year}{1981}.
\newblock \bibinfo{title}{{Thermal structure and dynamics of the Jovian
  atmosphere. I - The Great Red Spot}}.
\newblock \bibinfo{journal}{J. Geophys. Res.} \bibinfo{volume}{86},
  \bibinfo{pages}{8759--8767}.
\newblock \DOIprefix\doi{10.1029/JA086iA10p08759}.
%Type = Article
\bibitem[{{Flasar} et~al.(2004){Flasar}, {Kunde}, {Achterberg}, {Conrath},
  {Simon-Miller}, {Nixon}, {Gierasch}, {Romani}, {B{\'e}zard}, {Irwin},
  {Bjoraker}, {Brasunas}, {Jennings}, {Pearl}, {Smith}, {Orton}, {Spilker},
  {Carlson}, {Calcutt}, {Read}, {Taylor}, {Parrish}, {Barucci}, {Courtin},
  {Coustenis}, {Gautier}, {Lellouch}, {Marten}, {Prang{\'e}}, {Biraud},
  {Fouchet}, {Ferrari}, {Owen}, {Abbas}, {Samuelson}, {Raulin}, {Ade},
  {C{\'e}sarsky}, {Grossman} and {Coradini}}]{04flasar_jup}
\bibinfo{author}{{Flasar}, F.M.}, \bibinfo{author}{{Kunde}, V.G.},
  \bibinfo{author}{{Achterberg}, R.K.}, \bibinfo{author}{{Conrath}, B.J.},
  \bibinfo{author}{{Simon-Miller}, A.A.}, \bibinfo{author}{{Nixon}, C.A.},
  \bibinfo{author}{{Gierasch}, P.J.}, \bibinfo{author}{{Romani}, P.N.},
  \bibinfo{author}{{B{\'e}zard}, B.}, \bibinfo{author}{{Irwin}, P.},
  \bibinfo{author}{{Bjoraker}, G.L.}, \bibinfo{author}{{Brasunas}, J.C.},
  \bibinfo{author}{{Jennings}, D.E.}, \bibinfo{author}{{Pearl}, J.C.},
  \bibinfo{author}{{Smith}, M.D.}, \bibinfo{author}{{Orton}, G.S.},
  \bibinfo{author}{{Spilker}, L.J.}, \bibinfo{author}{{Carlson}, R.},
  \bibinfo{author}{{Calcutt}, S.B.}, \bibinfo{author}{{Read}, P.L.},
  \bibinfo{author}{{Taylor}, F.W.}, \bibinfo{author}{{Parrish}, P.},
  \bibinfo{author}{{Barucci}, A.}, \bibinfo{author}{{Courtin}, R.},
  \bibinfo{author}{{Coustenis}, A.}, \bibinfo{author}{{Gautier}, D.},
  \bibinfo{author}{{Lellouch}, E.}, \bibinfo{author}{{Marten}, A.},
  \bibinfo{author}{{Prang{\'e}}, R.}, \bibinfo{author}{{Biraud}, Y.},
  \bibinfo{author}{{Fouchet}, T.}, \bibinfo{author}{{Ferrari}, C.},
  \bibinfo{author}{{Owen}, T.C.}, \bibinfo{author}{{Abbas}, M.M.},
  \bibinfo{author}{{Samuelson}, R.E.}, \bibinfo{author}{{Raulin}, F.},
  \bibinfo{author}{{Ade}, P.}, \bibinfo{author}{{C{\'e}sarsky}, C.J.},
  \bibinfo{author}{{Grossman}, K.U.}, \bibinfo{author}{{Coradini}, A.},
  \bibinfo{year}{2004}.
\newblock \bibinfo{title}{{An intense stratospheric jet on Jupiter}}.
\newblock \bibinfo{journal}{Nature} \bibinfo{volume}{427},
  \bibinfo{pages}{132--135}.
%Type = Article
\bibitem[{{Fletcher} et~al.(2014){Fletcher}, {de Pater}, {Orton}, {Hammel},
  {Sitko} and {Irwin}}]{14fletcher_nep}
\bibinfo{author}{{Fletcher}, L.N.}, \bibinfo{author}{{de Pater}, I.},
  \bibinfo{author}{{Orton}, G.S.}, \bibinfo{author}{{Hammel}, H.B.},
  \bibinfo{author}{{Sitko}, M.L.}, \bibinfo{author}{{Irwin}, P.G.J.},
  \bibinfo{year}{2014}.
\newblock \bibinfo{title}{{Neptune at summer solstice: Zonal mean temperatures
  from ground-based observations, 2003-2007}}.
\newblock \bibinfo{journal}{Icarus} \bibinfo{volume}{231},
  \bibinfo{pages}{146--167}.
\newblock \DOIprefix\doi{10.1016/j.icarus.2013.11.035},
  \href{http://arxiv.org/abs/1311.7570}{{\tt arXiv:1311.7570}}.
%Type = Article
\bibitem[{{Fletcher} et~al.(2016a){Fletcher}, {Greathouse}, {Orton},
  {Sinclair}, {Giles}, {Irwin} and {Encrenaz}}]{16fletcher_texes}
\bibinfo{author}{{Fletcher}, L.N.}, \bibinfo{author}{{Greathouse}, T.K.},
  \bibinfo{author}{{Orton}, G.S.}, \bibinfo{author}{{Sinclair}, J.A.},
  \bibinfo{author}{{Giles}, R.S.}, \bibinfo{author}{{Irwin}, P.G.J.},
  \bibinfo{author}{{Encrenaz}, T.}, \bibinfo{year}{2016}a.
\newblock \bibinfo{title}{{Mid-infrared mapping of Jupiter's temperatures,
  aerosol opacity and chemical distributions with IRTF/TEXES}}.
\newblock \bibinfo{journal}{Icarus} \bibinfo{volume}{278},
  \bibinfo{pages}{128--161}.
\newblock \DOIprefix\doi{10.1016/j.icarus.2016.06.008},
  \href{http://arxiv.org/abs/1606.05498}{{\tt arXiv:1606.05498}}.
%Type = Article
\bibitem[{{Fletcher} et~al.(2016b){Fletcher}, {Irwin}, {Achterberg}, {Orton}
  and {Flasar}}]{16fletcher}
\bibinfo{author}{{Fletcher}, L.N.}, \bibinfo{author}{{Irwin}, P.G.J.},
  \bibinfo{author}{{Achterberg}, R.K.}, \bibinfo{author}{{Orton}, G.S.},
  \bibinfo{author}{{Flasar}, F.M.}, \bibinfo{year}{2016}b.
\newblock \bibinfo{title}{{Seasonal variability of Saturn's tropospheric
  temperatures, winds and para-H$_{2}$ from Cassini far-IR spectroscopy}}.
\newblock \bibinfo{journal}{Icarus} \bibinfo{volume}{264},
  \bibinfo{pages}{137--159}.
\newblock \DOIprefix\doi{10.1016/j.icarus.2015.09.009},
  \href{http://arxiv.org/abs/1509.02281}{{\tt arXiv:1509.02281}}.
%Type = Article
\bibitem[{{Fletcher} et~al.(2011){Fletcher}, {Orton}, {Rogers}, {Simon-Miller},
  {de Pater}, {Wong}, {Mousis}, {Irwin}, {Jacquesson} and
  {Yanamandra-Fisher}}]{11fletcher_fade}
\bibinfo{author}{{Fletcher}, L.N.}, \bibinfo{author}{{Orton}, G.S.},
  \bibinfo{author}{{Rogers}, J.H.}, \bibinfo{author}{{Simon-Miller}, A.A.},
  \bibinfo{author}{{de Pater}, I.}, \bibinfo{author}{{Wong}, M.H.},
  \bibinfo{author}{{Mousis}, O.}, \bibinfo{author}{{Irwin}, P.G.J.},
  \bibinfo{author}{{Jacquesson}, M.}, \bibinfo{author}{{Yanamandra-Fisher},
  P.A.}, \bibinfo{year}{2011}.
\newblock \bibinfo{title}{{Jovian temperature and cloud variability during the
  2009-2010 fade of the South Equatorial Belt}}.
\newblock \bibinfo{journal}{Icarus} \bibinfo{volume}{213},
  \bibinfo{pages}{564--580}.
\newblock \DOIprefix\doi{10.1016/j.icarus.2011.03.007}.
%Type = Article
\bibitem[{{Fletcher} et~al.(2009a){Fletcher}, {Orton}, {Teanby} and
  {Irwin}}]{09fletcher_ph3}
\bibinfo{author}{{Fletcher}, L.N.}, \bibinfo{author}{{Orton}, G.S.},
  \bibinfo{author}{{Teanby}, N.A.}, \bibinfo{author}{{Irwin}, P.G.J.},
  \bibinfo{year}{2009}a.
\newblock \bibinfo{title}{{Phosphine on Jupiter and Saturn from Cassini/CIRS}}.
\newblock \bibinfo{journal}{Icarus} \bibinfo{volume}{202},
  \bibinfo{pages}{543--564}.
\newblock \DOIprefix\doi{10.1016/j.icarus.2009.03.023}.
%Type = Article
\bibitem[{{Fletcher} et~al.(2009b){Fletcher}, {Orton}, {Yanamandra-Fisher},
  {Fisher}, {Parrish} and {Irwin}}]{09fletcher_imaging}
\bibinfo{author}{{Fletcher}, L.N.}, \bibinfo{author}{{Orton}, G.S.},
  \bibinfo{author}{{Yanamandra-Fisher}, P.}, \bibinfo{author}{{Fisher}, B.M.},
  \bibinfo{author}{{Parrish}, P.D.}, \bibinfo{author}{{Irwin}, P.G.J.},
  \bibinfo{year}{2009}b.
\newblock \bibinfo{title}{{Retrievals of atmospheric variables on the gas
  giants from ground-based mid-infrared imaging}}.
\newblock \bibinfo{journal}{Icarus} \bibinfo{volume}{200},
  \bibinfo{pages}{154--175}.
\newblock \DOIprefix\doi{10.1016/j.icarus.2008.11.019}.
%Type = Article
\bibitem[{{Fouchet} et~al.(2003){Fouchet}, {Lellouch} and
  {Feuchtgruber}}]{03fouchet}
\bibinfo{author}{{Fouchet}, T.}, \bibinfo{author}{{Lellouch}, E.},
  \bibinfo{author}{{Feuchtgruber}, H.}, \bibinfo{year}{2003}.
\newblock \bibinfo{title}{{The hydrogen ortho-to-para ratio in the
  stratospheres of the giant planets}}.
\newblock \bibinfo{journal}{Icarus} \bibinfo{volume}{161},
  \bibinfo{pages}{127--143}.
\newblock \DOIprefix\doi{10.1016/S0019-1035(02)00014-3}.
%Type = Article
\bibitem[{{Frommhold} et~al.(1984){Frommhold}, {Samuelson} and
  {Birnbaum}}]{84frommhold}
\bibinfo{author}{{Frommhold}, L.}, \bibinfo{author}{{Samuelson}, R.},
  \bibinfo{author}{{Birnbaum}, G.}, \bibinfo{year}{1984}.
\newblock \bibinfo{title}{{Hydrogen dimer structures in the far-infrared
  spectra of Jupiter and Saturn}}.
\newblock \bibinfo{journal}{ApJ Letters} \bibinfo{volume}{283},
  \bibinfo{pages}{L79--L82}.
\newblock \DOIprefix\doi{10.1086/184338}.
%Type = Article
\bibitem[{{Gautier} et~al.(1981){Gautier}, {Conrath}, {Flasar}, {Hanel},
  {Kunde}, {Chedin} and {Scott}}]{81gautier}
\bibinfo{author}{{Gautier}, D.}, \bibinfo{author}{{Conrath}, B.},
  \bibinfo{author}{{Flasar}, M.}, \bibinfo{author}{{Hanel}, R.},
  \bibinfo{author}{{Kunde}, V.}, \bibinfo{author}{{Chedin}, A.},
  \bibinfo{author}{{Scott}, N.}, \bibinfo{year}{1981}.
\newblock \bibinfo{title}{{The helium abundance of Jupiter from Voyager}}.
\newblock \bibinfo{journal}{Journal of Geophysical Research}
  \bibinfo{volume}{86}, \bibinfo{pages}{8713--8720}.
\newblock \DOIprefix\doi{10.1029/JA086iA10p08713}.
%Type = Article
\bibitem[{{Gehrz} et~al.(2009){Gehrz}, {Becklin}, {de Pater}, {Lester},
  {Roellig} and {Woodward}}]{09gehrz}
\bibinfo{author}{{Gehrz}, R.D.}, \bibinfo{author}{{Becklin}, E.E.},
  \bibinfo{author}{{de Pater}, I.}, \bibinfo{author}{{Lester}, D.F.},
  \bibinfo{author}{{Roellig}, T.L.}, \bibinfo{author}{{Woodward}, C.E.},
  \bibinfo{year}{2009}.
\newblock \bibinfo{title}{{A new window on the cosmos: The Stratospheric
  Observatory for Infrared Astronomy (SOFIA)}}.
\newblock \bibinfo{journal}{Advances in Space Research} \bibinfo{volume}{44},
  \bibinfo{pages}{413--432}.
\newblock \DOIprefix\doi{10.1016/j.asr.2009.04.011}.
%Type = Article
\bibitem[{{Gierasch} et~al.(1986){Gierasch}, {Magalhaes} and
  {Conrath}}]{86gierasch}
\bibinfo{author}{{Gierasch}, P.J.}, \bibinfo{author}{{Magalhaes}, J.A.},
  \bibinfo{author}{{Conrath}, B.J.}, \bibinfo{year}{1986}.
\newblock \bibinfo{title}{{Zonal mean properties of Jupiter's upper troposphere
  from Voyager infrared observations}}.
\newblock \bibinfo{journal}{Icarus} \bibinfo{volume}{67},
  \bibinfo{pages}{456--483}.
\newblock \DOIprefix\doi{10.1016/0019-1035(86)90125-9}.
%Type = Article
\bibitem[{Griffith et~al.(1992)Griffith, B{\'e}zard, Owen and
  Gautier}]{92griffith}
\bibinfo{author}{Griffith, C.}, \bibinfo{author}{B{\'e}zard, B.},
  \bibinfo{author}{Owen, T.}, \bibinfo{author}{Gautier, D.},
  \bibinfo{year}{1992}.
\newblock \bibinfo{title}{The tropospheric abundances of nh3 and ph3 in
  jupiter's great red spot, from voyager iris observations}.
\newblock \bibinfo{journal}{Icarus} \bibinfo{volume}{98},
  \bibinfo{pages}{82--93}.
%Type = Article
\bibitem[{{Hanel} et~al.(1982){Hanel}, {Conrath}, {Flasar}, {Kunde}, {Maguire},
  {Pearl}, {Pirraglia}, {Samuelson}, {Cruikshank}, {Gautier}, {Gierasch},
  {Horn} and {Ponnamperuma}}]{82hanel}
\bibinfo{author}{{Hanel}, R.}, \bibinfo{author}{{Conrath}, B.},
  \bibinfo{author}{{Flasar}, F.M.}, \bibinfo{author}{{Kunde}, V.},
  \bibinfo{author}{{Maguire}, W.}, \bibinfo{author}{{Pearl}, J.C.},
  \bibinfo{author}{{Pirraglia}, J.}, \bibinfo{author}{{Samuelson}, R.},
  \bibinfo{author}{{Cruikshank}, D.P.}, \bibinfo{author}{{Gautier}, D.},
  \bibinfo{author}{{Gierasch}, P.J.}, \bibinfo{author}{{Horn}, L.},
  \bibinfo{author}{{Ponnamperuma}, C.}, \bibinfo{year}{1982}.
\newblock \bibinfo{title}{{Infrared observations of the Saturnian system from
  Voyager 2}}.
\newblock \bibinfo{journal}{Science} \bibinfo{volume}{215},
  \bibinfo{pages}{544--548}.
%Type = Article
\bibitem[{{Hanel} et~al.(1979){Hanel}, {Conrath}, {Flasar}, {Kunde}, {Lowman},
  {Maguire}, {Pearl}, {Pirraglia}, {Samuelson}, {Gautier}, {Gierasch}, {Kumar}
  and {Ponnamperuma}}]{79hanel}
\bibinfo{author}{{Hanel}, R.}, \bibinfo{author}{{Conrath}, B.},
  \bibinfo{author}{{Flasar}, M.}, \bibinfo{author}{{Kunde}, V.},
  \bibinfo{author}{{Lowman}, P.}, \bibinfo{author}{{Maguire}, W.},
  \bibinfo{author}{{Pearl}, J.}, \bibinfo{author}{{Pirraglia}, J.},
  \bibinfo{author}{{Samuelson}, R.}, \bibinfo{author}{{Gautier}, D.},
  \bibinfo{author}{{Gierasch}, P.}, \bibinfo{author}{{Kumar}, S.},
  \bibinfo{author}{{Ponnamperuma}, C.}, \bibinfo{year}{1979}.
\newblock \bibinfo{title}{{Infrared observations of the Jovian system from
  Voyager 1}}.
\newblock \bibinfo{journal}{Science} \bibinfo{volume}{204},
  \bibinfo{pages}{972--976}.
\newblock \DOIprefix\doi{10.1126/science.204.4396.972-a}.
%Type = Article
\bibitem[{{Hanel} et~al.(1977){Hanel}, {Conrath}, {Kunde}, {Lowman}, {Maguire},
  {Pearl}, {Pirraglia}, {Gautier}, {Gierasch} and {Kumar}}]{77hanel}
\bibinfo{author}{{Hanel}, R.}, \bibinfo{author}{{Conrath}, B.},
  \bibinfo{author}{{Kunde}, V.}, \bibinfo{author}{{Lowman}, P.},
  \bibinfo{author}{{Maguire}, W.}, \bibinfo{author}{{Pearl}, J.},
  \bibinfo{author}{{Pirraglia}, J.}, \bibinfo{author}{{Gautier}, D.},
  \bibinfo{author}{{Gierasch}, P.}, \bibinfo{author}{{Kumar}, S.},
  \bibinfo{year}{1977}.
\newblock \bibinfo{title}{{The Voyager infrared spectroscopy and radiometry
  investigation}}.
\newblock \bibinfo{journal}{Space Science Reviews} \bibinfo{volume}{21},
  \bibinfo{pages}{129--157}.
\newblock \DOIprefix\doi{10.1007/BF00200848}.
%Type = Article
\bibitem[{{Herter} et~al.(2012){Herter}, {Adams}, {De Buizer}, {Gull},
  {Schoenwald}, {Henderson}, {Keller}, {Nikola}, {Stacey} and
  {Vacca}}]{12herter}
\bibinfo{author}{{Herter}, T.L.}, \bibinfo{author}{{Adams}, J.D.},
  \bibinfo{author}{{De Buizer}, J.M.}, \bibinfo{author}{{Gull}, G.E.},
  \bibinfo{author}{{Schoenwald}, J.}, \bibinfo{author}{{Henderson}, C.P.},
  \bibinfo{author}{{Keller}, L.D.}, \bibinfo{author}{{Nikola}, T.},
  \bibinfo{author}{{Stacey}, G.}, \bibinfo{author}{{Vacca}, W.D.},
  \bibinfo{year}{2012}.
\newblock \bibinfo{title}{{First Science Observations with SOFIA/FORCAST: The
  FORCAST Mid-infrared Camera}}.
\newblock \bibinfo{journal}{ApJ Letters} \bibinfo{volume}{749},
  \bibinfo{pages}{L18}.
\newblock \DOIprefix\doi{10.1088/2041-8205/749/2/L18},
  \href{http://arxiv.org/abs/1202.5021}{{\tt arXiv:1202.5021}}.
%Type = Article
\bibitem[{{Herter} et~al.(2013){Herter}, {Vacca}, {Adams}, {Keller},
  {Schoenwald}, {Hirsch}, {Wang}, {De Buizer}, {Helton} and
  {Llorens}}]{13herter}
\bibinfo{author}{{Herter}, T.L.}, \bibinfo{author}{{Vacca}, W.D.},
  \bibinfo{author}{{Adams}, J.D.}, \bibinfo{author}{{Keller}, L.D.},
  \bibinfo{author}{{Schoenwald}, J.}, \bibinfo{author}{{Hirsch}, L.},
  \bibinfo{author}{{Wang}, J.}, \bibinfo{author}{{De Buizer}, J.M.},
  \bibinfo{author}{{Helton}, L.A.}, \bibinfo{author}{{Llorens}, M.C.},
  \bibinfo{year}{2013}.
\newblock \bibinfo{title}{{Data Reduction and Early Science Calibration for
  FORCAST, A Mid-Infrared Camera for SOFIA}}.
\newblock \bibinfo{journal}{PASP} \bibinfo{volume}{125},
  \bibinfo{pages}{1393--1404}.
\newblock \DOIprefix\doi{10.1086/674144}.
%Type = Article
\bibitem[{Irwin et~al.(2008)Irwin, Teanby, de~Kok, Fletcher, Howett, Tsang,
  Wilson, Calcutt, Nixon and Parrish}]{08irwin}
\bibinfo{author}{Irwin, P.}, \bibinfo{author}{Teanby, N.},
  \bibinfo{author}{de~Kok, R.}, \bibinfo{author}{Fletcher, L.},
  \bibinfo{author}{Howett, C.}, \bibinfo{author}{Tsang, C.},
  \bibinfo{author}{Wilson, C.}, \bibinfo{author}{Calcutt, S.},
  \bibinfo{author}{Nixon, C.}, \bibinfo{author}{Parrish, P.},
  \bibinfo{year}{2008}.
\newblock \bibinfo{title}{{The NEMESIS planetary atmosphere radiative transfer
  and retrieval tool}}.
\newblock \bibinfo{journal}{Journal of Quantitative Spectroscopy and Radiative
  Transfer} \bibinfo{volume}{109}, \bibinfo{pages}{1136--1150}.
%Type = Article
\bibitem[{{Irwin} et~al.(2004){Irwin}, {Parrish}, {Fouchet}, {Calcutt},
  {Taylor}, {Simon-Miller} and {Nixon}}]{04irwin}
\bibinfo{author}{{Irwin}, P.G.J.}, \bibinfo{author}{{Parrish}, P.},
  \bibinfo{author}{{Fouchet}, T.}, \bibinfo{author}{{Calcutt}, S.B.},
  \bibinfo{author}{{Taylor}, F.W.}, \bibinfo{author}{{Simon-Miller}, A.A.},
  \bibinfo{author}{{Nixon}, C.A.}, \bibinfo{year}{2004}.
\newblock \bibinfo{title}{{Retrievals of jovian tropospheric phosphine from
  Cassini/CIRS}}.
\newblock \bibinfo{journal}{Icarus} \bibinfo{volume}{172},
  \bibinfo{pages}{37--49}.
\newblock \DOIprefix\doi{10.1016/j.icarus.2003.09.027}.
%Type = Inproceedings
\bibitem[{{Keller} et~al.(2010){Keller}, {Deen}, {Jaffe}, {Ennico}, {Greene},
  {Adams}, {Herter} and {Sloan}}]{10keller}
\bibinfo{author}{{Keller}, L.}, \bibinfo{author}{{Deen}, C.P.},
  \bibinfo{author}{{Jaffe}, D.T.}, \bibinfo{author}{{Ennico}, K.A.},
  \bibinfo{author}{{Greene}, T.P.}, \bibinfo{author}{{Adams}, J.D.},
  \bibinfo{author}{{Herter}, T.}, \bibinfo{author}{{Sloan}, G.C.},
  \bibinfo{year}{2010}.
\newblock \bibinfo{title}{{Progress report on FORCAST grism spectroscopy as a
  future general observer instrument mode on SOFIA}}, in:
  \bibinfo{booktitle}{Ground-based and Airborne Instrumentation for Astronomy
  III}, p. \bibinfo{pages}{77356N}.
\newblock \DOIprefix\doi{10.1117/12.857127}.
%Type = Article
\bibitem[{{Kessler} et~al.(1996){Kessler}, {Steinz}, {Anderegg}, {Clavel},
  {Drechsel}, {Estaria}, {Faelker}, {Riedinger}, {Robson}, {Taylor} and
  {Xim{\'e}nez de Ferr{\'a}n}}]{96kessler}
\bibinfo{author}{{Kessler}, M.F.}, \bibinfo{author}{{Steinz}, J.A.},
  \bibinfo{author}{{Anderegg}, M.E.}, \bibinfo{author}{{Clavel}, J.},
  \bibinfo{author}{{Drechsel}, G.}, \bibinfo{author}{{Estaria}, P.},
  \bibinfo{author}{{Faelker}, J.}, \bibinfo{author}{{Riedinger}, J.R.},
  \bibinfo{author}{{Robson}, A.}, \bibinfo{author}{{Taylor}, B.G.},
  \bibinfo{author}{{Xim{\'e}nez de Ferr{\'a}n}, S.}, \bibinfo{year}{1996}.
\newblock \bibinfo{title}{{The Infrared Space Observatory (ISO) mission.}}
\newblock \bibinfo{journal}{Astron. Astrophys} \bibinfo{volume}{315},
  \bibinfo{pages}{L27--L31}.
%Type = Article
\bibitem[{{Kostiuk} et~al.(1993){Kostiuk}, {Romani}, {Espenak} and
  {Livengood}}]{93kostiuk}
\bibinfo{author}{{Kostiuk}, T.}, \bibinfo{author}{{Romani}, P.},
  \bibinfo{author}{{Espenak}, F.}, \bibinfo{author}{{Livengood}, T.A.},
  \bibinfo{year}{1993}.
\newblock \bibinfo{title}{{Temperature and abundances in the Jovian auroral
  stratosphere. 2: Ethylene as a probe of the microbar region}}.
\newblock \bibinfo{journal}{Journal of Geophysical Research}
  \bibinfo{volume}{98}, \bibinfo{pages}{18}.
\newblock \DOIprefix\doi{10.1029/93JE01332}.
%Type = Article
\bibitem[{{Leovy} et~al.(1991){Leovy}, {Friedson} and {Orton}}]{91leovy}
\bibinfo{author}{{Leovy}, C.B.}, \bibinfo{author}{{Friedson}, A.J.},
  \bibinfo{author}{{Orton}, G.S.}, \bibinfo{year}{1991}.
\newblock \bibinfo{title}{{The quasiquadrennial oscillation of Jupiter's
  equatorial stratosphere}}.
\newblock \bibinfo{journal}{Nature} \bibinfo{volume}{354},
  \bibinfo{pages}{380--382}.
\newblock \DOIprefix\doi{10.1038/354380a0}.
%Type = Article
\bibitem[{{Livengood} et~al.(1993){Livengood}, {Kostiuk} and
  {Espenak}}]{93livengood}
\bibinfo{author}{{Livengood}, T.A.}, \bibinfo{author}{{Kostiuk}, T.},
  \bibinfo{author}{{Espenak}, F.}, \bibinfo{year}{1993}.
\newblock \bibinfo{title}{{Temperature and abundances in the Jovian auroral
  stratosphere. 1: Ethane as a probe of the millibar region}}.
\newblock \bibinfo{journal}{Journal of Geophysical Research}
  \bibinfo{volume}{98}, \bibinfo{pages}{18}.
\newblock \DOIprefix\doi{10.1029/93JE01043}.
%Type = Article
\bibitem[{{Martonchik} et~al.(1984){Martonchik}, {Orton} and
  {Appleby}}]{84martonchik}
\bibinfo{author}{{Martonchik}, J.V.}, \bibinfo{author}{{Orton}, G.S.},
  \bibinfo{author}{{Appleby}, J.F.}, \bibinfo{year}{1984}.
\newblock \bibinfo{title}{{Optical properties of NH3 ice from the far infrared
  to the near ultraviolet}}.
\newblock \bibinfo{journal}{Applied Optics} \bibinfo{volume}{23},
  \bibinfo{pages}{541--547}.
%Type = Article
\bibitem[{{Massie} and {Hunten}(1982)}]{82massie}
\bibinfo{author}{{Massie}, S.T.}, \bibinfo{author}{{Hunten}, D.M.},
  \bibinfo{year}{1982}.
\newblock \bibinfo{title}{{Conversion of para and ortho hydrogen in the Jovian
  planets}}.
\newblock \bibinfo{journal}{Icarus} \bibinfo{volume}{49},
  \bibinfo{pages}{213--226}.
\newblock \DOIprefix\doi{10.1016/0019-1035(82)90073-2}.
%Type = Article
\bibitem[{Matcheva et~al.(2005)Matcheva, Conrath, Gierasch and
  Flasar}]{05matcheva}
\bibinfo{author}{Matcheva, K.}, \bibinfo{author}{Conrath, B.},
  \bibinfo{author}{Gierasch, P.}, \bibinfo{author}{Flasar, F.},
  \bibinfo{year}{2005}.
\newblock \bibinfo{title}{{The cloud structure of the jovian atmosphere as seen
  by the Cassini/CIRS experiment}}.
\newblock \bibinfo{journal}{Icarus} \bibinfo{volume}{179},
  \bibinfo{pages}{432--448}.
%Type = Article
\bibitem[{{Meyer} et~al.(1989a){Meyer}, {Borysow} and
  {Frommhold}}]{89meyer_fund}
\bibinfo{author}{{Meyer}, W.}, \bibinfo{author}{{Borysow}, A.},
  \bibinfo{author}{{Frommhold}, L.}, \bibinfo{year}{1989}a.
\newblock \bibinfo{title}{{Absorption spectra of H2-H2 pairs in the fundamental
  band}}.
\newblock \bibinfo{journal}{Physical Review A} \bibinfo{volume}{40},
  \bibinfo{pages}{6931--6949}.
\newblock \DOIprefix\doi{10.1103/PhysRevA.40.6931}.
%Type = Article
\bibitem[{{Meyer} et~al.(1989b){Meyer}, {Frommhold} and {Birnbaum}}]{89meyer}
\bibinfo{author}{{Meyer}, W.}, \bibinfo{author}{{Frommhold}, L.},
  \bibinfo{author}{{Birnbaum}, G.}, \bibinfo{year}{1989}b.
\newblock \bibinfo{title}{{Rototranslational absorption spectra of H2-H2 pairs
  in the far infrared}}.
\newblock \bibinfo{journal}{Physical Review A} \bibinfo{volume}{39},
  \bibinfo{pages}{2434--2448}.
\newblock \DOIprefix\doi{10.1103/PhysRevA.39.2434}.
%Type = Article
\bibitem[{{Minnaert}(1941)}]{41minnaert}
\bibinfo{author}{{Minnaert}, M.}, \bibinfo{year}{1941}.
\newblock \bibinfo{title}{{The reciprocity principle in lunar photometry}}.
\newblock \bibinfo{journal}{{Astrophys. J.}} \bibinfo{volume}{93},
  \bibinfo{pages}{403--410}.
\newblock \DOIprefix\doi{10.1086/144279}.
%Type = Article
\bibitem[{{Niemann} et~al.(1998){Niemann}, {Atreya}, {Carignan}, {Donahue},
  {Haberman}, {Harpold}, {Hartle}, {Hunten}, {Kasprzak}, {Mahaffy}, {Owen} and
  {Way}}]{98niemann}
\bibinfo{author}{{Niemann}, H.B.}, \bibinfo{author}{{Atreya}, S.K.},
  \bibinfo{author}{{Carignan}, G.R.}, \bibinfo{author}{{Donahue}, T.M.},
  \bibinfo{author}{{Haberman}, J.A.}, \bibinfo{author}{{Harpold}, D.N.},
  \bibinfo{author}{{Hartle}, R.E.}, \bibinfo{author}{{Hunten}, D.M.},
  \bibinfo{author}{{Kasprzak}, W.T.}, \bibinfo{author}{{Mahaffy}, P.R.},
  \bibinfo{author}{{Owen}, T.C.}, \bibinfo{author}{{Way}, S.H.},
  \bibinfo{year}{1998}.
\newblock \bibinfo{title}{{The composition of the Jovian atmosphere as
  determined by the Galileo probe mass spectrometer}}.
\newblock \bibinfo{journal}{J. Geophys. Res.} \bibinfo{volume}{103},
  \bibinfo{pages}{22831--22846}.
\newblock \DOIprefix\doi{10.1029/98JE01050}.
%Type = Article
\bibitem[{{Nixon} et~al.(2010){Nixon}, {Achterberg}, {Romani}, {Allen},
  {Zhang}, {Teanby}, {Irwin} and {Flasar}}]{10nixon}
\bibinfo{author}{{Nixon}, C.A.}, \bibinfo{author}{{Achterberg}, R.K.},
  \bibinfo{author}{{Romani}, P.N.}, \bibinfo{author}{{Allen}, M.},
  \bibinfo{author}{{Zhang}, X.}, \bibinfo{author}{{Teanby}, N.A.},
  \bibinfo{author}{{Irwin}, P.G.J.}, \bibinfo{author}{{Flasar}, F.M.},
  \bibinfo{year}{2010}.
\newblock \bibinfo{title}{{Abundances of Jupiter's trace hydrocarbons from
  Voyager and Cassini}}.
\newblock \bibinfo{journal}{Plan. \& Space Sci.} \bibinfo{volume}{58},
  \bibinfo{pages}{1667--1680}.
\newblock \DOIprefix\doi{10.1016/j.pss.2010.05.008},
  \href{http://arxiv.org/abs/1005.3959}{{\tt arXiv:1005.3959}}.
%Type = Article
\bibitem[{{Norwood} et~al.(2016){Norwood}, {Moses}, {Fletcher}, {Orton},
  {Irwin}, {Atreya}, {Rages}, {Cavali{\'e}}, {S{\'a}nchez-Lavega}, {Hueso} and
  {Chanover}}]{16norwood}
\bibinfo{author}{{Norwood}, J.}, \bibinfo{author}{{Moses}, J.},
  \bibinfo{author}{{Fletcher}, L.N.}, \bibinfo{author}{{Orton}, G.},
  \bibinfo{author}{{Irwin}, P.G.J.}, \bibinfo{author}{{Atreya}, S.},
  \bibinfo{author}{{Rages}, K.}, \bibinfo{author}{{Cavali{\'e}}, T.},
  \bibinfo{author}{{S{\'a}nchez-Lavega}, A.}, \bibinfo{author}{{Hueso}, R.},
  \bibinfo{author}{{Chanover}, N.}, \bibinfo{year}{2016}.
\newblock \bibinfo{title}{{Giant Planet Observations with the James Webb Space
  Telescope}}.
\newblock \bibinfo{journal}{PASP} \bibinfo{volume}{128},
  \bibinfo{pages}{018005}.
\newblock \DOIprefix\doi{10.1088/1538-3873/128/959/018005},
  \href{http://arxiv.org/abs/1510.06205}{{\tt arXiv:1510.06205}}.
%Type = Article
\bibitem[{{Orton} et~al.(2015){Orton}, {Fletcher}, {Encrenaz}, {Leyrat}, {Roe},
  {Fujiyoshi} and {Pantin}}]{15orton}
\bibinfo{author}{{Orton}, G.S.}, \bibinfo{author}{{Fletcher}, L.N.},
  \bibinfo{author}{{Encrenaz}, T.}, \bibinfo{author}{{Leyrat}, C.},
  \bibinfo{author}{{Roe}, H.G.}, \bibinfo{author}{{Fujiyoshi}, T.},
  \bibinfo{author}{{Pantin}, E.}, \bibinfo{year}{2015}.
\newblock \bibinfo{title}{{Thermal imaging of Uranus: Upper-tropospheric
  temperatures one season after Voyager}}.
\newblock \bibinfo{journal}{Icarus} \bibinfo{volume}{260},
  \bibinfo{pages}{94--102}.
\newblock \DOIprefix\doi{10.1016/j.icarus.2015.07.004}.
%Type = Article
\bibitem[{{Orton} et~al.(2014){Orton}, {Fletcher}, {Moses}, {Mainzer}, {Hines},
  {Hammel}, {Martin-Torres}, {Burgdorf}, {Merlet} and {Line}}]{14orton}
\bibinfo{author}{{Orton}, G.S.}, \bibinfo{author}{{Fletcher}, L.N.},
  \bibinfo{author}{{Moses}, J.I.}, \bibinfo{author}{{Mainzer}, A.K.},
  \bibinfo{author}{{Hines}, D.}, \bibinfo{author}{{Hammel}, H.B.},
  \bibinfo{author}{{Martin-Torres}, F.J.}, \bibinfo{author}{{Burgdorf}, M.},
  \bibinfo{author}{{Merlet}, C.}, \bibinfo{author}{{Line}, M.R.},
  \bibinfo{year}{2014}.
\newblock \bibinfo{title}{{Mid-infrared spectroscopy of Uranus from the Spitzer
  Infrared Spectrometer: 1. Determination of the mean temperature structure of
  the upper troposphere and stratosphere}}.
\newblock \bibinfo{journal}{Icarus} \bibinfo{volume}{243},
  \bibinfo{pages}{494--513}.
\newblock \DOIprefix\doi{10.1016/j.icarus.2014.07.010},
  \href{http://arxiv.org/abs/1407.2120}{{\tt arXiv:1407.2120}}.
%Type = Article
\bibitem[{{Orton} et~al.(1991){Orton}, {Friedson}, {Caldwell}, {Hammel},
  {Baines}, {Bergstralh}, {Martin}, {Malcom}, {West}, {Golisch}, {Griep},
  {Kaminski}, {Tokunaga}, {Baron} and {Shure}}]{91orton}
\bibinfo{author}{{Orton}, G.S.}, \bibinfo{author}{{Friedson}, A.J.},
  \bibinfo{author}{{Caldwell}, J.}, \bibinfo{author}{{Hammel}, H.B.},
  \bibinfo{author}{{Baines}, K.H.}, \bibinfo{author}{{Bergstralh}, J.T.},
  \bibinfo{author}{{Martin}, T.Z.}, \bibinfo{author}{{Malcom}, M.E.},
  \bibinfo{author}{{West}, R.A.}, \bibinfo{author}{{Golisch}, W.F.},
  \bibinfo{author}{{Griep}, D.M.}, \bibinfo{author}{{Kaminski}, C.D.},
  \bibinfo{author}{{Tokunaga}, A.T.}, \bibinfo{author}{{Baron}, R.},
  \bibinfo{author}{{Shure}, M.}, \bibinfo{year}{1991}.
\newblock \bibinfo{title}{{Thermal maps of Jupiter - Spatial organization and
  time dependence of stratospheric temperatures, 1980 to 1990}}.
\newblock \bibinfo{journal}{Science} \bibinfo{volume}{252},
  \bibinfo{pages}{537--542}.
%Type = Article
\bibitem[{{Orton} et~al.(1994){Orton}, {Friedson}, {Yanamandra-Fisher},
  {Caldwell}, {Hammel}, {Baines}, {Bergstralh}, {Martin}, {West}, {Veeder},
  {Lynch}, {Russell}, {Malcom}, {Golisch}, {Griep}, {Kaminski}, {Tokunaga},
  {Herbst} and {Shure}}]{94orton}
\bibinfo{author}{{Orton}, G.S.}, \bibinfo{author}{{Friedson}, A.J.},
  \bibinfo{author}{{Yanamandra-Fisher}, P.A.}, \bibinfo{author}{{Caldwell},
  J.}, \bibinfo{author}{{Hammel}, H.B.}, \bibinfo{author}{{Baines}, K.H.},
  \bibinfo{author}{{Bergstralh}, J.T.}, \bibinfo{author}{{Martin}, T.Z.},
  \bibinfo{author}{{West}, R.A.}, \bibinfo{author}{{Veeder}, Jr., G.J.},
  \bibinfo{author}{{Lynch}, D.K.}, \bibinfo{author}{{Russell}, R.},
  \bibinfo{author}{{Malcom}, M.E.}, \bibinfo{author}{{Golisch}, W.F.},
  \bibinfo{author}{{Griep}, D.M.}, \bibinfo{author}{{Kaminski}, C.D.},
  \bibinfo{author}{{Tokunaga}, A.T.}, \bibinfo{author}{{Herbst}, T.},
  \bibinfo{author}{{Shure}, M.}, \bibinfo{year}{1994}.
\newblock \bibinfo{title}{{Spatial Organization and Time Dependence of
  Jupiter's Tropospheric Temperatures, 1980-1993}}.
\newblock \bibinfo{journal}{Science} \bibinfo{volume}{265},
  \bibinfo{pages}{625--631}.
\newblock \DOIprefix\doi{10.1126/science.265.5172.625}.
%Type = Article
\bibitem[{{Orton} et~al.(2007){Orton}, {Gustafsson}, {Burgdorf} and
  {Meadows}}]{07orton}
\bibinfo{author}{{Orton}, G.S.}, \bibinfo{author}{{Gustafsson}, M.},
  \bibinfo{author}{{Burgdorf}, M.}, \bibinfo{author}{{Meadows}, V.},
  \bibinfo{year}{2007}.
\newblock \bibinfo{title}{{Revised Ab Initio Models for H$_2$-H$_2$ Collision
  Induced Absorption at Low Temperatures}}.
\newblock \bibinfo{journal}{Icarus} \bibinfo{volume}{189},
  \bibinfo{pages}{544--549}.
%Type = Article
\bibitem[{Read et~al.(2006)Read, Gierasch and Conrath}]{06read_grs}
\bibinfo{author}{Read, P.}, \bibinfo{author}{Gierasch, P.},
  \bibinfo{author}{Conrath, B.}, \bibinfo{year}{2006}.
\newblock \bibinfo{title}{{Mapping potential-vorticity dynamics on Jupiter. II:
  The Great Red Spot from Voyager 1 and 2 data}}.
\newblock \bibinfo{journal}{Quarterly Journal of the Royal Meteorological
  Society} \bibinfo{volume}{132}, \bibinfo{pages}{1605--1625}.
%Type = Article
\bibitem[{{Richard} et~al.(2012){Richard}, {Gordon}, {Rothman}, {Abel},
  {Frommhold}, {Gustafsson}, {Hartmann}, {Hermans}, {Lafferty}, {Orton},
  {Smith} and {Tran}}]{12richard}
\bibinfo{author}{{Richard}, C.}, \bibinfo{author}{{Gordon}, I.E.},
  \bibinfo{author}{{Rothman}, L.S.}, \bibinfo{author}{{Abel}, M.},
  \bibinfo{author}{{Frommhold}, L.}, \bibinfo{author}{{Gustafsson}, M.},
  \bibinfo{author}{{Hartmann}, J.M.}, \bibinfo{author}{{Hermans}, C.},
  \bibinfo{author}{{Lafferty}, W.J.}, \bibinfo{author}{{Orton}, G.S.},
  \bibinfo{author}{{Smith}, K.M.}, \bibinfo{author}{{Tran}, H.},
  \bibinfo{year}{2012}.
\newblock \bibinfo{title}{{New section of the HITRAN database:
  Collision-induced absorption (CIA)}}.
\newblock \bibinfo{journal}{Journal of Quantitative Spectroscopy and Radiative
  Transfer} \bibinfo{volume}{113}, \bibinfo{pages}{1276--1285}.
\newblock \DOIprefix\doi{10.1016/j.jqsrt.2011.11.004}.
%Type = Book
\bibitem[{{Rodgers}(2000)}]{00rodgers}
\bibinfo{author}{{Rodgers}, C.D.}, \bibinfo{year}{2000}.
\newblock \bibinfo{title}{{Inverse Methods for Atmospheric Remote Sounding:
  Theory and Practice}}.
\newblock \bibinfo{publisher}{World Scientific}.
%Type = Article
\bibitem[{{Sada} et~al.(1996){Sada}, {Beebe} and {Conrath}}]{96sada}
\bibinfo{author}{{Sada}, P.V.}, \bibinfo{author}{{Beebe}, R.F.},
  \bibinfo{author}{{Conrath}, B.J.}, \bibinfo{year}{1996}.
\newblock \bibinfo{title}{{Comparison of the Structure and Dynamics of
  Jupiter's Great Red SPOT between the Voyager 1 and 2 Encounters}}.
\newblock \bibinfo{journal}{Icarus} \bibinfo{volume}{119},
  \bibinfo{pages}{311--335}.
\newblock \DOIprefix\doi{10.1006/icar.1996.0022}.
%Type = Article
\bibitem[{{Schaefer} and {McKellar}(1990)}]{90schaefer}
\bibinfo{author}{{Schaefer}, J.}, \bibinfo{author}{{McKellar}, A.R.W.},
  \bibinfo{year}{1990}.
\newblock \bibinfo{title}{{Faint features of the rotational S$_{0}$(0) and
  S$_{0}$(1) transitions of H$_{2}$. A comparison of calculations and
  measurements at 77 K}}.
\newblock \bibinfo{journal}{Zeitschrift fur Physik D Atoms Molecules Clusters}
  \bibinfo{volume}{15}, \bibinfo{pages}{51--65}.
\newblock \DOIprefix\doi{10.1007/BF01436911}.
%Type = Article
\bibitem[{{Seiff} et~al.(1998){Seiff}, {Kirk}, {Knight}, {Young}, {Mihalov},
  {Young}, {Milos}, {Schubert}, {Blanchard} and {Atkinson}}]{98seiff}
\bibinfo{author}{{Seiff}, A.}, \bibinfo{author}{{Kirk}, D.B.},
  \bibinfo{author}{{Knight}, T.C.D.}, \bibinfo{author}{{Young}, R.E.},
  \bibinfo{author}{{Mihalov}, J.D.}, \bibinfo{author}{{Young}, L.A.},
  \bibinfo{author}{{Milos}, F.S.}, \bibinfo{author}{{Schubert}, G.},
  \bibinfo{author}{{Blanchard}, R.C.}, \bibinfo{author}{{Atkinson}, D.},
  \bibinfo{year}{1998}.
\newblock \bibinfo{title}{{Thermal structure of Jupiter's atmosphere near the
  edge of a 5-{$\mu$}m hot spot in the north equatorial belt}}.
\newblock \bibinfo{journal}{J. Geophys. Res.} \bibinfo{volume}{103},
  \bibinfo{pages}{22857--22890}.
\newblock \DOIprefix\doi{10.1029/98JE01766}.
%Type = Article
\bibitem[{Simon-Miller et~al.(2002)Simon-Miller, Gierasch, Beebe, Conrath,
  Flasar and Achterberg}]{02simon}
\bibinfo{author}{Simon-Miller, A.}, \bibinfo{author}{Gierasch, P.},
  \bibinfo{author}{Beebe, R.}, \bibinfo{author}{Conrath, B.},
  \bibinfo{author}{Flasar, F.}, \bibinfo{author}{Achterberg, R.},
  \bibinfo{year}{2002}.
\newblock \bibinfo{title}{{New Observational Results Concerning Jupiter's Great
  Red Spot}}.
\newblock \bibinfo{journal}{Icarus} \bibinfo{volume}{158},
  \bibinfo{pages}{249--266}.
%Type = Article
\bibitem[{{Simon-Miller} et~al.(2000){Simon-Miller}, {Conrath}, {Gierasch} and
  {Beebe}}]{00simon}
\bibinfo{author}{{Simon-Miller}, A.A.}, \bibinfo{author}{{Conrath}, B.},
  \bibinfo{author}{{Gierasch}, P.J.}, \bibinfo{author}{{Beebe}, R.F.},
  \bibinfo{year}{2000}.
\newblock \bibinfo{title}{{A detection of water ice on Jupiter with Voyager
  IRIS}}.
\newblock \bibinfo{journal}{Icarus} \bibinfo{volume}{145},
  \bibinfo{pages}{454--461}.
\newblock \DOIprefix\doi{10.1006/icar.2000.6359}.
%Type = Article
\bibitem[{{Simon-Miller} et~al.(2006){Simon-Miller}, {Conrath}, {Gierasch},
  {Orton}, {Achterberg}, {Flasar} and {Fisher}}]{06simon}
\bibinfo{author}{{Simon-Miller}, A.A.}, \bibinfo{author}{{Conrath}, B.J.},
  \bibinfo{author}{{Gierasch}, P.J.}, \bibinfo{author}{{Orton}, G.S.},
  \bibinfo{author}{{Achterberg}, R.K.}, \bibinfo{author}{{Flasar}, F.M.},
  \bibinfo{author}{{Fisher}, B.M.}, \bibinfo{year}{2006}.
\newblock \bibinfo{title}{{Jupiter's atmospheric temperatures: From Voyager
  IRIS to Cassini CIRS}}.
\newblock \bibinfo{journal}{Icarus} \bibinfo{volume}{180},
  \bibinfo{pages}{98--112}.
\newblock \DOIprefix\doi{10.1016/j.icarus.2005.07.019}.
%Type = Article
\bibitem[{{Sinclair} et~al.(2014){Sinclair}, {Irwin}, {Fletcher}, {Greathouse},
  {Guerlet}, {Hurley} and {Merlet}}]{14sinclair}
\bibinfo{author}{{Sinclair}, J.A.}, \bibinfo{author}{{Irwin}, P.G.J.},
  \bibinfo{author}{{Fletcher}, L.N.}, \bibinfo{author}{{Greathouse}, T.},
  \bibinfo{author}{{Guerlet}, S.}, \bibinfo{author}{{Hurley}, J.},
  \bibinfo{author}{{Merlet}, C.}, \bibinfo{year}{2014}.
\newblock \bibinfo{title}{{From Voyager-IRIS to Cassini-CIRS: Interannual
  variability in Saturn's stratosphere?}}
\newblock \bibinfo{journal}{Icarus} \bibinfo{volume}{233},
  \bibinfo{pages}{281--292}.
\newblock \DOIprefix\doi{10.1016/j.icarus.2014.02.009}.
%Type = Article
\bibitem[{{Wang} et~al.(2015){Wang}, {Gierasch}, {Lunine} and
  {Mousis}}]{15wang}
\bibinfo{author}{{Wang}, D.}, \bibinfo{author}{{Gierasch}, P.J.},
  \bibinfo{author}{{Lunine}, J.I.}, \bibinfo{author}{{Mousis}, O.},
  \bibinfo{year}{2015}.
\newblock \bibinfo{title}{{New insights on Jupiter's deep water abundance from
  disequilibrium species}}.
\newblock \bibinfo{journal}{Icarus} \bibinfo{volume}{250},
  \bibinfo{pages}{154--164}.
\newblock \DOIprefix\doi{10.1016/j.icarus.2014.11.026},
  \href{http://arxiv.org/abs/1412.0690}{{\tt arXiv:1412.0690}}.
%Type = Article
\bibitem[{{Wong} et~al.(2004){Wong}, {Bjoraker}, {Smith}, {Flasar} and
  {Nixon}}]{04wong}
\bibinfo{author}{{Wong}, M.H.}, \bibinfo{author}{{Bjoraker}, G.L.},
  \bibinfo{author}{{Smith}, M.D.}, \bibinfo{author}{{Flasar}, F.M.},
  \bibinfo{author}{{Nixon}, C.A.}, \bibinfo{year}{2004}.
\newblock \bibinfo{title}{{Identification of the 10-{$\mu$}m ammonia ice
  feature on Jupiter}}.
\newblock \bibinfo{journal}{Plan. \& Space Sci.} \bibinfo{volume}{52},
  \bibinfo{pages}{385--395}.
\newblock \DOIprefix\doi{10.1016/j.pss.2003.06.005}.
%Type = Article
\bibitem[{{Young} et~al.(2012){Young}, {Becklin}, {Marcum}, {Roellig}, {De
  Buizer}, {Herter}, {G{\"u}sten}, {Dunham}, {Temi}, {Andersson}, {Backman},
  {Burgdorf}, {Caroff}, {Casey}, {Davidson}, {Erickson}, {Gehrz}, {Harper},
  {Harvey}, {Helton}, {Horner}, {Howard}, {Klein}, {Krabbe}, {McLean}, {Meyer},
  {Miles}, {Morris}, {Reach}, {Rho}, {Richter}, {Roeser}, {Sandell}, {Sankrit},
  {Savage}, {Smith}, {Shuping}, {Vacca}, {Vaillancourt}, {Wolf} and
  {Zinnecker}}]{12young}
\bibinfo{author}{{Young}, E.T.}, \bibinfo{author}{{Becklin}, E.E.},
  \bibinfo{author}{{Marcum}, P.M.}, \bibinfo{author}{{Roellig}, T.L.},
  \bibinfo{author}{{De Buizer}, J.M.}, \bibinfo{author}{{Herter}, T.L.},
  \bibinfo{author}{{G{\"u}sten}, R.}, \bibinfo{author}{{Dunham}, E.W.},
  \bibinfo{author}{{Temi}, P.}, \bibinfo{author}{{Andersson}, B.G.},
  \bibinfo{author}{{Backman}, D.}, \bibinfo{author}{{Burgdorf}, M.},
  \bibinfo{author}{{Caroff}, L.J.}, \bibinfo{author}{{Casey}, S.C.},
  \bibinfo{author}{{Davidson}, J.A.}, \bibinfo{author}{{Erickson}, E.F.},
  \bibinfo{author}{{Gehrz}, R.D.}, \bibinfo{author}{{Harper}, D.A.},
  \bibinfo{author}{{Harvey}, P.M.}, \bibinfo{author}{{Helton}, L.A.},
  \bibinfo{author}{{Horner}, S.D.}, \bibinfo{author}{{Howard}, C.D.},
  \bibinfo{author}{{Klein}, R.}, \bibinfo{author}{{Krabbe}, A.},
  \bibinfo{author}{{McLean}, I.S.}, \bibinfo{author}{{Meyer}, A.W.},
  \bibinfo{author}{{Miles}, J.W.}, \bibinfo{author}{{Morris}, M.R.},
  \bibinfo{author}{{Reach}, W.T.}, \bibinfo{author}{{Rho}, J.},
  \bibinfo{author}{{Richter}, M.J.}, \bibinfo{author}{{Roeser}, H.P.},
  \bibinfo{author}{{Sandell}, G.}, \bibinfo{author}{{Sankrit}, R.},
  \bibinfo{author}{{Savage}, M.L.}, \bibinfo{author}{{Smith}, E.C.},
  \bibinfo{author}{{Shuping}, R.Y.}, \bibinfo{author}{{Vacca}, W.D.},
  \bibinfo{author}{{Vaillancourt}, J.E.}, \bibinfo{author}{{Wolf}, J.},
  \bibinfo{author}{{Zinnecker}, H.}, \bibinfo{year}{2012}.
\newblock \bibinfo{title}{{Early Science with SOFIA, the Stratospheric
  Observatory For Infrared Astronomy}}.
\newblock \bibinfo{journal}{ApJ Letters} \bibinfo{volume}{749},
  \bibinfo{pages}{L17}.
\newblock \DOIprefix\doi{10.1088/2041-8205/749/2/L17},
  \href{http://arxiv.org/abs/1205.0791}{{\tt arXiv:1205.0791}}.
%Type = Article
\bibitem[{{Zhang} et~al.(2013){Zhang}, {West}, {Banfield} and
  {Yung}}]{13zhang_aer}
\bibinfo{author}{{Zhang}, X.}, \bibinfo{author}{{West}, R.A.},
  \bibinfo{author}{{Banfield}, D.}, \bibinfo{author}{{Yung}, Y.L.},
  \bibinfo{year}{2013}.
\newblock \bibinfo{title}{{Stratospheric aerosols on Jupiter from Cassini
  observations}}.
\newblock \bibinfo{journal}{Icarus} \bibinfo{volume}{226},
  \bibinfo{pages}{159--171}.
\newblock \DOIprefix\doi{10.1016/j.icarus.2013.05.020}.

\end{thebibliography}




%% \bibitem must have one of the following forms:
%%   \bibitem[Jones et al.(1990)]{key}...
%%   \bibitem[Jones et al.(1990)Jones, Baker, and Williams]{key}...
%%   \bibitem[Jones et al., 1990]{key}...
%%   \bibitem[\protect\citeauthoryear{Jones, Baker, and Williams}{Jones
%%       et al.}{1990}]{key}...
%%   \bibitem[\protect\citeauthoryear{Jones et al.}{1990}]{key}...
%%   \bibitem[\protect\astroncite{Jones et al.}{1990}]{key}...
%%   \bibitem[\protect\citename{Jones et al., }1990]{key}...
%%   \harvarditem[Jones et al.]{Jones, Baker, and Williams}{1990}{key}...
%%


% \end{thebibliography}

\end{document}